# Crystallization of piezoceramic films on glass via flash lamp annealing

Longfei Song,[1,2,†] Juliette Cardoletti,[1,†] Alfredo Blázquez Martínez,[1,2] Andreja Benčan,[3] Brigita Kmet,[3] Stéphanie Girod,[1] Emmanuel Defay,[1] Sebastjan Glinšek[1,*]

[1] Materials Research and Technology Department, Luxembourg Institute of Science and Technology, 41 rue du Brill, L-4422 Belvaux, Luxembourg

[2] University of Luxembourg, 41 rue du Brill, L-4422 Belvaux, Luxembourg

[3] Electronic Ceramics Department, Jožef Stefan Institute, Jamova cesta 39, 1000 Ljubljana, Slovenia

[†] These authors contributed equally: L. S. and J. C.

*Corresponding author: sebastjan.glinsek@list.lu

**Abstract**

Integration of thin-film oxide piezoelectrics on glass is imperative for the next generation of transparent electronics to attain sensing and actuating functions. However, their crystallization temperature (above 650 °C) is incompatible with most glasses. We developed a flash lamp process for growth of piezoelectric lead zirconate titanate films. The process enables crystallization on various types of glasses in a few seconds only. Functional properties of these films are comparable to the films processed with standard rapid thermal annealing at 700 °C. A surface haptic device was fabricated with a 1 µm-thick film (piezoelectric $e_{33,f}$ of -5 C m$^{-2}$). Its ultrasonic surface deflection reached 1.5 µm at 60 V, sufficient for its use in surface rendering applications. This flash lamp annealing process is compatible with large glass sheets and roll-to-roll processing and has the potential to significantly expand the applications of piezoelectric devices on glass.

# 1. Introduction

An important trend in next-generation large-scale electronics is the integration of functional films on glass wafers for smart windows and display screens. Among the available materials for sensors and actuators, piezoelectric oxide thin films are outstanding because of their superior electromechanical response compared to nitrides and polymers [1–5]. The two last ones are excellent materials for resonators and energy harvesters, respectively, and can be processed at low temperatures (e.g. <350 °C for AlN - and ~150 °C for polyvinyl difluoride (PVDF)-based materials). But much lower piezoelectric coefficients prevent them to replace perovskites, especially for actuator applications [6–8]. The key for their successful integration of perovskites in microelectromechanical systems (MEMS) has been efficient processing, which allows for the preparation of high-quality films in a controllable and reproducible manner. Chemical solution deposition (CSD) is among the most popular fabrication methods due to its low cost, flexibility in chemical composition and compatibility with large-scale microelectronics [9,10]. The method is continuously evolving and emerging digital printing and roll-to-roll processing technologies will enable high-speed, high-throughput and large-scale additive manufacturing [5,11].

In CSD processing the as-deposited amorphous phase is transformed into a crystalline perovskite piezoelectric phase via high-temperature annealing. The crystallisation process traditionally relies on the use of isothermal heating in tube, box, or rapid thermal annealing (RTA) furnaces (see Figure 1a) [12]. The typical processing temperatures are above 650 °C and the total annealing time (including heating and cooling) is in the order of tens of minutes. As the strain point - the maximum temperature at which glass can be used without experiencing creep - of most commercial glasses lies below 600 °C, the development of low-temperature processing is required [13,14].

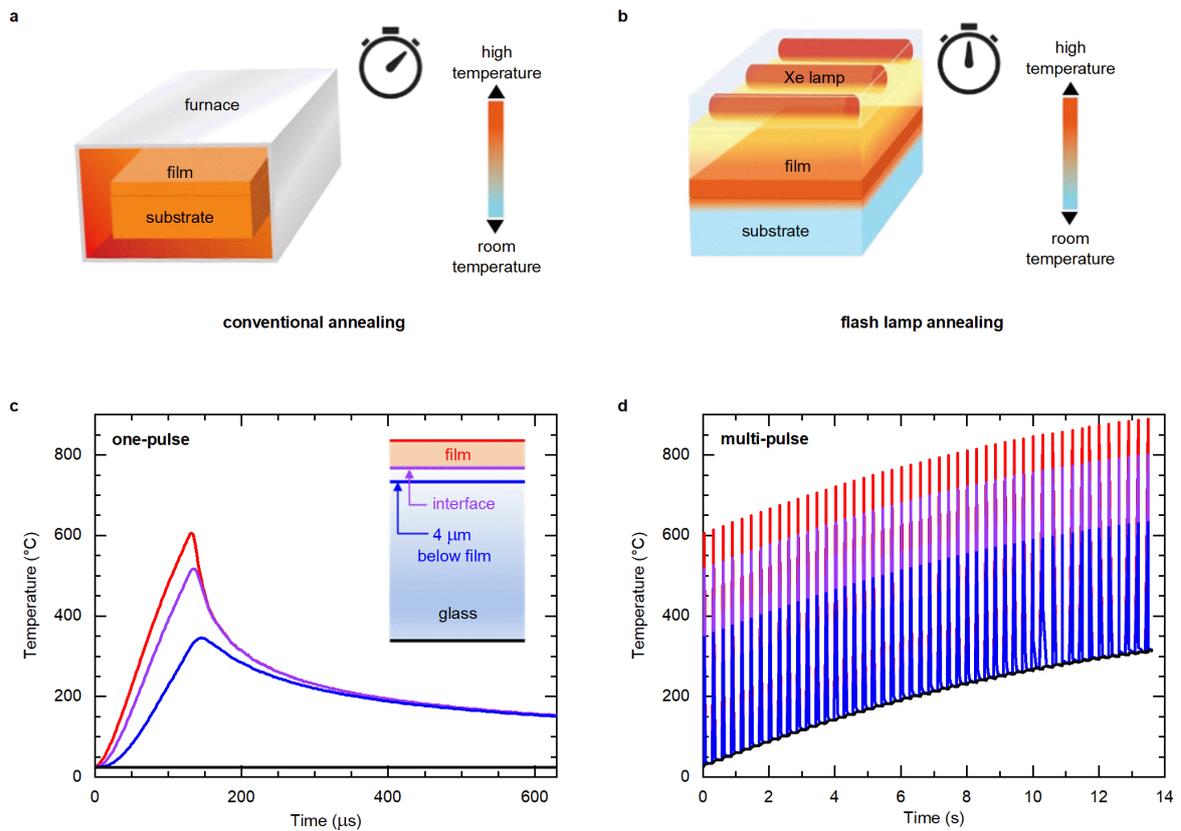

Figure 1. **Schematic representation of annealing processes.** Heat distribution in a film, its substrate, and processing time in the case of **a** conventional box furnace annealing and **b** flash lamp annealing. **Finite element modelling**. Temperature profiles during **c** single-pulse and **d** 50-pulse flash lamp annealing of a 170 nm-PZT/fused silica glass stack. The red, purple, blue, and black lines correspond to the temperature profiles of the top PZT surface, the interface between the film and glass, 4 μm below this interface, and the bottom of the substrate (500 μm), respectively. Absolute temperature values are indicative only. The schematic structure of the sample is shown in the inset of **c**. The pulse duration, energy density and repetition rate are 130 μs, 3 J cm$^{-2}$, and 3.5 Hz, respectively.

Efforts have been dedicated to lowering processing temperature, including by photochemical processing [15,16], annealing under high-pressure of $O_2/O_3$ [17], combustion synthesis [18], and laser annealing [19,20]. While the global temperature has been lowered, the price to pay is either long annealing times (often several hours) or the use of reactive atmosphere, which limit the use of these processes in large-scale and high-throughput production. Although laser annealing allows for the fast crystallization of films, the small laser spot size (typically in the µm²-mm² range) imposes the need for raster scanning of the laser beam, leading to inhomogeneities in the films [21]. Pulsed laser deposition (PLD) method can lead to good films at 445 °C, however, low pressures (0.05 mbar) are required, and the method is difficult for scale-up [22,23]. The indirect integration on glass via transfer process has been successful [24], however, it significantly complicates the process and adds processing steps.

Flash lamp annealing (FLA), i.e., annealing through the absorption of sub-millisecond light pulses with broad spectrum and strong intensity, enables selective annealing of the films on temperature-sensitive substrates [25] and is schematically compared to box-furnace annealing in Figure 1b. FLA can be performed in ambient atmosphere in a few seconds and the irradiated areas can reach several hundreds of cm². These features are compatible with high-throughput, low-cost and large-scale roll-to-roll production [26]. FLA has been successfully used for processing different functional thin films, such as conductors (e.g., indium tin oxide, metal nanoparticle films) [27,28], semiconductors (e.g., indium zinc oxide) [29,30], dielectrics (e.g., alumina, zirconia) [31,32], hybrid perovskites for absorbers in solar cells [33] and light-emitting diodes [34]. Yao et al. induced nano-crystallization of piezoelectric lead zirconate titanate on glass and polyimide substrates, however, macroscopic electromechanical properties, which are imperative for sensor and actuator applications, were not demonstrated [35].

Piezoelectrics have been recently demonstrated as efficient actuators for haptics. The technology is based on ultrasonically vibrating surface, which can modulate forces at the interface between a finger and a vibrating plate. The so-called friction-modulation effect is applicable in touchscreens with foreseen applications such as displays for automotive industry, displays for visually disabled people, or sliders for control of heating/cooling [36]. The effect depends strongly on the out-of-plane displacement, in-plane wavelength, and frequency of the standing acoustic wave, which is created in the screen. For technology commercialization, the following criteria have to be fulfilled: amplitude must be larger than 1 μm (enabling detection with a human finger and significant decrease of a friction coefficient) [37,38], wavelength must be below ~15 mm (enabling detection with a human finger) [39] and its frequency should be beyond 25 kHz (enabling silent operation) [40].

Inspired by the above results and guided by finite element modelling (FEM), we developed a process that enables macroscopic crystallization of solution processed PbZr$_{0.53}$Ti$_{0.47}$O$_3$ (PZT) films on a wide variety of glasses. First, we demonstrate a fast process (several seconds per crystallization) on 1 μm-thick PZT thin films on fused silica, resulting in remanent polarization $P_r$, dielectric permittivity $\varepsilon_r$, dielectric losses tan$\delta$ and piezoelectric coefficient $e_{33,f}$ values of 12 μC cm$^{-2}$, 450, 5 %, and -5 C m$^{-2}$, respectively. A surface haptic device was realized on alumina-borosilicate glass (AF32, Schott), i.e., a standard substrate for semiconductor and MEMS industries. The device is based on a 1 μm-thick PZT film and its ultrasonic-range surface deflection reaches 1.5 μm at 60 V, which fulfils requirement of a deflection of 1 μm for its commercialization in texture rendering operation [37]. Finally, we demonstrate the universality of FLA process by direct growth of PZT films on soda-lime, i.e., the most widespread type of glass.

Hence, in this paper, we have shown that it is possible to manufacture a functional device (a haptic transducer) based on piezoelectric perovskite thin films deposited on a glass substrate and sintered with a specific Flash Lamp Annealing process. Moreover, we showed that a distinctive feature of the latter is that this crystallization can be performed on glass substrates that cannot withstand temperature larger than 400°C.

## 2. Results

*2.1. Processes Design and Finite Element Modelling*

Results of the finite element modelling are presented in Figures 1c and 1d, where a 170 nm PZT/fused silica glass configuration was modelled. The absorbance of an amorphous PZT layer (~28.5 %) was estimated with a bolometer placed below the sample, considering that glass is mainly transparent in the ~300-1000 nm wavelength range of the Xe lamp utilised in this work (see Supplementary Note 1 for further information). The modelling of a 130 µs pulse with an energy density of 3 J cm$^{-2}$, indicates that the temperature at the top surface of PZT should reach 600 °C, which is sufficiently high to crystallize PZT into the piezoelectric perovskite phase [9]. Through the glass thickness, the temperature drops significantly as the process is non-adiabatic and performed in ambient environment. This maintains the interface between the film and the substrate at relatively low temperature, i.e., the temperature exceeds 400 °C for only 50 µs per pulse (see the purple line in Figure 1c). 4 µm below this interface the temperature remains below 350 °C. To provide enough energy and time for complete crystallization to occur, multi-pulse annealing is necessary. 30 pulses at 3.5 Hz enable the temperature of PZT to reach 800 °C while the bulk of the glass substrates remains below 400 °C (Figure 1d).

*2.2. Growth, Phase Composition and Microstructure*

The films were deposited through a standard CSD process using 2-methoxyethanol as solvent, lead acetate and transition metals alkoxides precursors [10]. The solutions were spin coated, dried at 130 °C and pyrolyzed at 350 °C on hot plates. The deposition-drying-pyrolysis process was repeated four times to achieve a thickness of 170 nm. The crystallization was done with flash lamp annealing. Thick films were obtained by repeating the whole process several times.

Films on fused silica were investigated first and their phase composition was analysed as a function of the number of FLA light pulses. Pulsing was performed with the same conditions as described in the previous section and in Supplementary Note 2. After 10 pulses, {100} and {110} reflections of the perovskite phase appear in the grazing incidence X-ray diffraction pattern (Figure 2a) of 170 nm-thick films. Reflection of the non-piezoelectric pyrochlore phase (a small peak at $2\theta = 29°$), which is kinetically stabilized during annealing [41], disappears when the number of pulses is increased to 30. When the number of pulses reaches 100, the intensity of perovskite reflections increases indicating enhanced crystallinity. The phase evolution agrees with the temperature evolution predicted with FEM during multi-pulse annealing (Figure 1d), i.e., gradual increase of temperature stabilizes the perovskite phase. The standard $\theta$-$2\theta$ X-ray diffraction (XRD) pattern of 1 μm-thick PZT on fused silica is shown in Supplementary Figure 2, and only perovskite peaks are present. In contrast, the perovskite phase does not form in the film annealed at 700 °C in RTA (Supplementary Figure 3) [42]. To the best of our knowledge, without the use of a nucleation layer (or bottom electrode), successful crystallization of solution-processed PZT films on amorphous substrates has not yet been reported, showing that the developed FLA process is even excelling a typical layer-by-layer deposition of solution-processed films in standard furnaces. It also enables the growth of films in a controllable and repeatable manner.

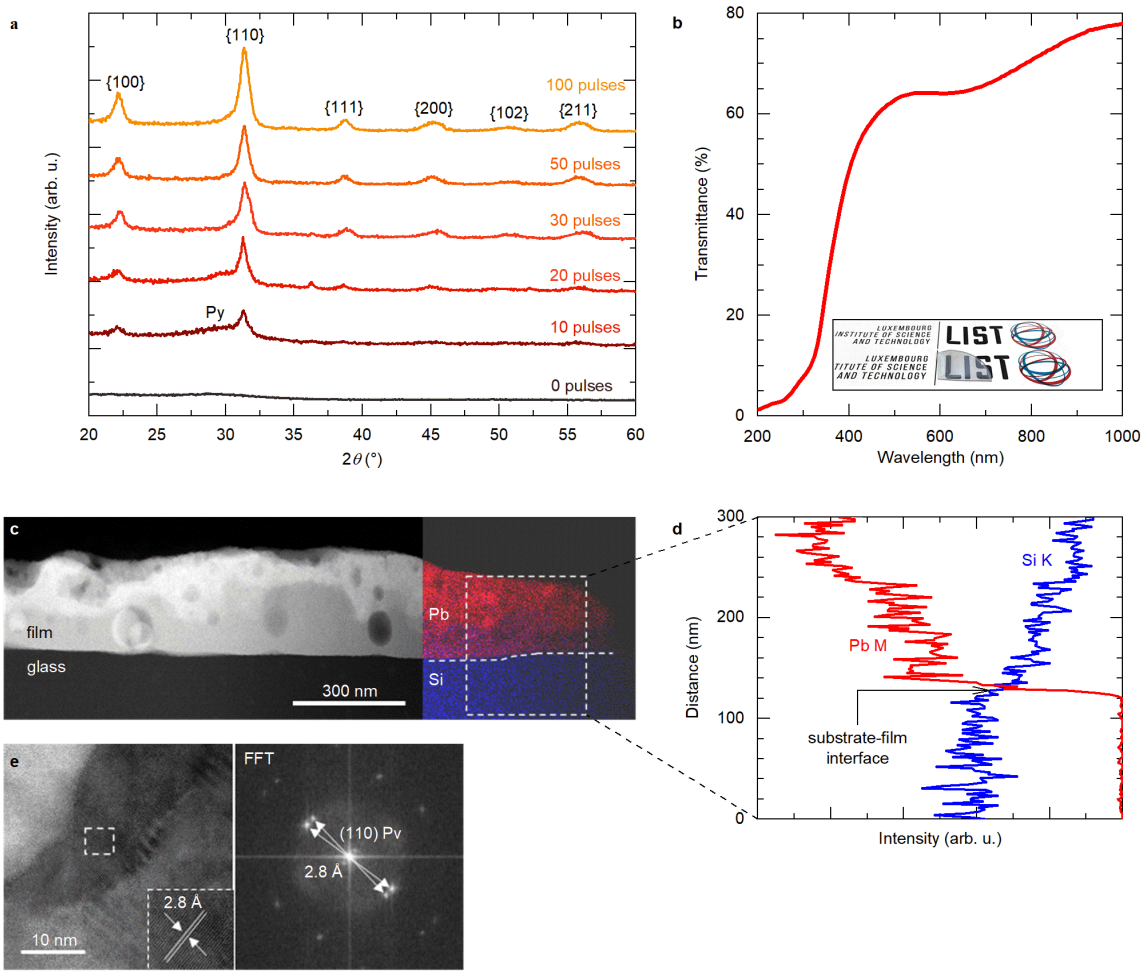

Figure 2. **Microstructural and optical characterization of 170 nm-thick PZT films on fused silica glass. a** GIXRD patterns of films annealed with different number of light pulses. In **a** perovskite reflections are denoted according to a powder diffraction file (PDF) No 01-070-4264 [43]. Py indicates the pyrochlore reflection according to PDF No 04-014-5162 [43]. **b** Transmittance of the 50-pulse film. Inset shows its optical appearance. STEM results of the 100-pulse film: **c** cross sectional dark-field STEM image and energy-dispersive X-ray spectroscopy system (EDS) analysis across substrate/film interface (marked by dashed line) using Pb M and Si K line, **d** EDS analysis across substrate-film interface using the Pb M and Si K lines showing partial diffusion of Si into the PZT film. The dashed area in **c** marks the region where the **d** EDS line analysis was performed, **e** high-resolution bright-field STEM image of two perovskite grains with corresponding FFT image showing (110) planes. Flash lamp annealing was performed with energy density, pulse duration and repetition rate of 3 J cm$^{-2}$, 130 μs, and 3.5 Hz, respectively.

Transmittance of the 170 nm-thick films (50 pulses) at a wavelength of 550 nm is 64 %, as shown in Figure 2b, together with its visual appearance. Transparency is preserved in the 1 μm film on 2'' glass wafer (inset of Supplementary Figure 4).

A cross-sectional scanning transmission electron microscope (STEM) analysis was also performed. A dark field STEM image (Figure 2c) reveals relatively porous granular microstructure, which agrees with the scanning electron microscopy micrograph of the 1 μm thick film shown in Supplementary Figure 5. The high-resolution STEM image, with the corresponding Fast Fourier Transform (FFT) (inset in Figure 2e), shows a (110) plane reflection with an interplanar spacing of 2.8 Å, which is further demonstrated in the zoomed-in STEM image (inset in Figure 2e). This interplanar spacing value is in good agreement with the perovskite phase of PZT and is additional confirmation of its presence in the film. While the film is chemically inhomogeneous (Supplementary Figure 6) the energy dispersive X-ray spectroscopy 2D map (Figure 2c) shows no diffusion of Pb into the substrate (as commonly observed in conventionally processed films) [44]. It is also observed that there was partial diffusion of silicon into the PZT film, at a depth of approximately 100 nm (Figure 2d). However, this diffusion did not result in the formation of secondary phase that would compromise the ferroelectric and piezoelectric properties of the PZT film, as confirmed earlier in grazing incidence X-ray diffraction (GIXRD) patterns and later by electrical measurements.

*2.3. Electromechanical Characterization*

This section is focused on the films prepared with 50 pulses, as they show optimal ferroelectric response. Electrical properties were measured in interdigitated geometry employing surface Pt electrodes. Polarization versus electric field *P*(**E**) loops are initially pinched (Supplementary Figure 7). This is most likely caused by the presence of charged defects (such as oxygen vacancies) in the films, which at low electric fields pin ferroelectric domain walls [45]. Electric-field cycling (wake-up) enables their redistribution, which unpins the polarization. Indeed, the pinched hysteresis opens during cycling (Supplementary Figure 7). Different number of wake-up cycles in different films could be due to different concentration of defects in the films and/or their different pinning energy.

Ferroelectric and piezoelectric properties are shown in Figure 3 and were obtained on a 1 µm-thick film on fused silica glass after $10^4$ wake-up cycles. The maximum polarization $P_{max}$ and remanent polarization $P_r$ are 25 µC cm$^{-2}$ and 12 µC cm$^{-2}$, respectively. Its coercive field $E_c$ is 68 kV cm$^{-1}$. Two sharp peaks, which are linked to domain switching, are observed in the current loop (Figure 3a). Similar results were obtained on 170 nm- and 500 nm-thick films (10 and 11 µC cm$^{-2}$ in $P_r$, respectively, see Supplementary Figures 7 and 8). The displacement of a cantilever structure shows a typical butterfly loop (Figure 3b). At 100 V the vertical displacement at the free end of the cantilever is 700 nm, corresponding to a piezoelectric coefficient $e_{33,f}$ of -5 C m$^{-2}$ [46].

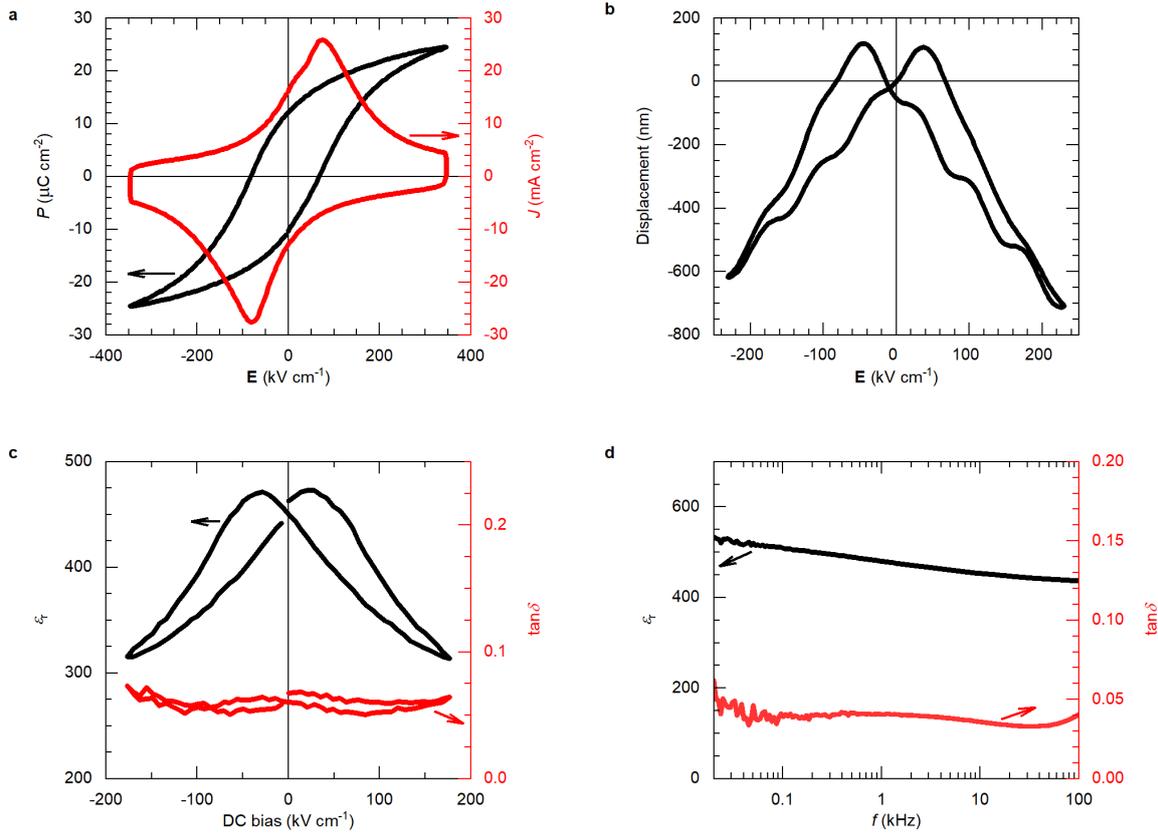

Figure 3. **Electromechanical characterization of a 1 μm-thick PZT film on fused silica glass**. **a** Ferroelectric and **b** displacement characterizations of the PZT film at 100 Hz and 11 Hz, respectively. Dielectric measurements of the PZT film: relative permittivity $\varepsilon_r$ and dielectric losses $\tan\delta$ as function of **c** DC bias at 1 kHz and **d** frequency $f$. The samples were processed with 50 pulses per layer with energy density, pulse duration and repetition rate of 3 J cm$^{-2}$, 130 μs, and 3.5 Hz, respectively.

Butterfly loops were also observed in electric-field dependence of relative permittivity $\varepsilon_r$ and dielectric losses $\tan\delta$ of the 1 μm-thick film (Figure 3c). Their zero-field values are 450 and 5 %, respectively. These quantities were also measured as functions of the small signal frequency (Figure 3d). While permittivity slightly decreases with increasing frequency, losses remain practically constant, which are typical signatures of perovskite ferroelectric films [47]. To have better overview of the results, a table with ferroelectric, piezoelectric and optical properties of the 170 nm, 500 nm and 1 μm-thick films on fused silica substrates is provided in Supplementary Table 1.

*2.4. Device Application*

To demonstrate suitability of the flash lamp annealing process for piezoelectric applications, we fabricated and characterized a surface haptic device [48]. AF32 glass, which is utilized in semiconductor and MEMS industry, was used [49]. A 1 μm-thick film was employed to increase the force of the actuator during its operation (see Supplementary Note 3.2.1). Phase composition and electrical properties of the film are comparable to the films on fused silica, as shown in Supplementary Note 3.

As schematically shown in Figure 4a, the dimensions of the fabricated haptic device are 15.4×3 mm$^2$. The thickness of the glass and film are 0.3 mm and 1 μm, respectively. Two actuating areas were created by fabricating 100 nm-thick Pt interdigitated (IDE) electrodes with a gap of 3 μm between the fingers. The distance between these two actuators is 8.4 mm. More details about the haptic device fabrication and measurements are described in the Methods and the Supplementary Note 3.2. The out-of-plane displacement at one of the resonance antinodes is shown as a function of frequency in Figure 4b. At 40.2 kHz the device reveals a peak in deflection, which corresponds to its mechanical resonance at anti-symmetric ($A_0$) Lamb mode. For further info on Lamb mode analysis see Supplementary Note 3.2.3 and Ref. [39]. Its position is at the same value as predicted by finite element modelling, whose details are described in Supplementary Note 3.2.2.

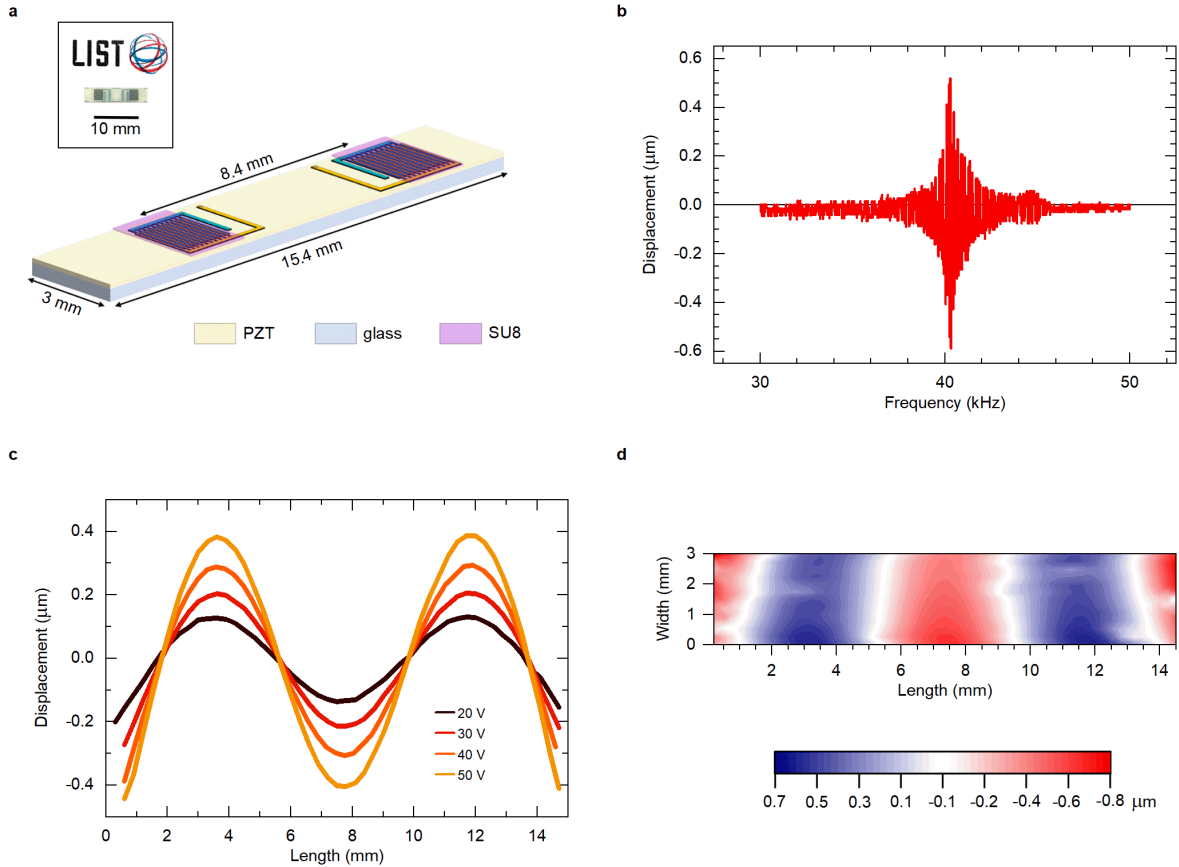

Figure 4. **Electromechanical characterization of the haptic device. a** Schematic of the device made of the flash lamp annealed 1 µm-thick PZT on AF32 glass. Two sets of interdigitated Pt electrodes are placed on top to create the actuators. Inset shows its visual appearance. **b** Out-of-plane displacement of the surface measured as a function of frequency at one of the antinodes. 60 $V_{pp}$ (30 $V_{AC}$+30 $V_{DC}$) were applied. **c** The displacement along *x*-axis (length of the device) excited at various driving voltages. **d** 2D map of the displacement at 60 $V_{pp}$. **c** and **d** are measured at the resonance frequency of 40.2 kHz.

The out-of-plane surface displacement in the *x*-direction (along the length of the device) is shown in Figure 4c. It was measured with an excitation voltage varying from 20 to 50 $V_{pp}$ at the resonance frequency of 40.2 kHz. Four nodes equally spaced along its length are observed, which is in line with the wave-shape obtained from the modelling (Supplementary Figure 11). Figure 4d shows a 2D displacement map of the haptic plate excited at 60 $V_{pp}$ (30 $V_{AC}$ + 30 $V_{DC}$) at the resonance. The device exhibits a maximum peak-to-peak deflection of 1.5 µm, beyond the specification of 1 μm, which can be detected by a human finger [37]. It also confirms a neat stationary Lamb wave. $e_{33,f}$ can be extracted by matching experimental deflection value with modelling [50] and the obtained value is -4.5 C m$^{-2}$, which is well in-line with the value obtained via cantilever measurements on fused silica (Figure 3b).

Performance of the device is compared to the previously reported piezoelectric haptic devices on glass substrates based on spin-coated [42] and inkjet-printed [5] PZT films with interdigitated geometry in Supplementary Table 2. Present device consumes 35 mW, which is comparable to the other two devices. Note also that 60 V can be applied to handheld devices by using either a cascade of several application-specific integrated circuits (ASICs) that can increase the output voltage from 3.3 to 100 V or by using an inductor L to make an LC resonator at the resonant frequency [42].

### 2.5. Flash Lamp Annealing for Films on Soda Lime Glass

To generalize the approach, we also used this flash lamp process to crystallize PZT films on soda-lime glass, which can only sustain temperatures below 400 °C. Because of the lower thermal conductivity of soda lime (1.0 W m$^{-1}$ K$^{-1}$) compared to fused silica (1.4 W m$^{-1}$ K$^{-1}$), the rate of heat transfer is slower, resulting in a higher temperature at the interface between the film and glass. A two-step process, resembling nucleation and growth in CSD process [9], was developed. In the first step, 6 pulses with higher power density of 14.7 kW cm$^{-2}$ (2.5 J cm$^{-2}$ and 170 μs) are applied. In the second step, 240 pulses with a lower power density of 10.0 kW cm$^{-2}$ (2.5 J cm$^{-2}$ and 250 μs) are applied to grow the film at a lower temperature, thereby preventing the occurrence of cracks. For both steps, the repetition rate is set to 0.5 Hz to increase the heat diffusion through the sample. This process is described in Supplementary Note 4, where also more details on phase composition and electrical properties are described (Supplementary

Figures 13 and 14). The films on soda lime glass also crystallize in the perovskite phase and have $P_r$ of 8 μC cm$^{-2}$, $\varepsilon_r$ of 400, and tan$\delta$ of 2 %, respectively.

## 3. Discussion

Macroscopic ferroelectric and piezoelectric properties of the films crystallized during flash lamp annealing are demonstrated in this work. The previously unreported process of two phases where either nucleation or grain growth of the perovskite phase is dominating has been enabled by the tool design, which enables creation of high-power pulses (above 20 kW cm$^{-2}$) with high repetition rates (3 Hz) [35,51–53]. Detailed comparison with the literature can be found in Supplementary Note 5.

To better understand the effectiveness of flash lamp annealing process we, conducted a comparison of the electrical properties between 1 μm-thick flash lamp-processed films and RTA-processed films on fused silica glass [42]. Table 1 displays the corresponding electrical parameters. It is worth noting that, in the RTA process, a hafnia buffer layer was applied to impede Pb diffusion, and that lead titanate oxide (PTO) nucleation was implemented to facilitate growth along {100} and therefore enhance the piezoelectric response [54]. Despite the doubling of the $\varepsilon_r$ value enabled by PTO nucleation, the ferroelectric and piezoelectric properties of the material remain relatively comparable.

Table 1. **Comparison of ferroelectric, dielectric and piezoelectric properties of chemical solution deposited PZT films processed by FLA, RTA and LA.** In Ref. [42] a lead titanate oxide nucleation layer was used to promote growth along (100) orientation, which enhances piezoelectric response [54]. MIM stands for metal-insulator-metal structure.

| Process | Substrate | Structure | $P_r$ (μC cm$^{-2}$) | $\varepsilon_r$ | $\tan\delta$ | $e_{33,f}$ (C m$^{-2}$) | Ref. |
|---|---|---|---|---|---|---|---|
| RTA | Fused silica | IDE | 16 | 900 | 0.05 | -8 | Glinsek et al. [42] |
| FLA | Fused silica | IDE | 11 | 450 | 0.05 | -5 | This work |
| FLA | AF32 | IDE | 5 | 350 | 0.07 | -5 | This work |
| LA | Platinised silicon | MIM | 28 | - | - | -4 | Fink et al. [20] |

It is also interesting to compare flash lamp process with other low-temperature processes of chemical solution-deposited PZT films. While reduced processing temperature has often been achieved, macroscopic piezoelectric properties of such films are seldomly reported, as demonstrated in a recent review paper [12]. The highest value reported so far in low-temperature films is $d_{33,f}$ = 49 pm V$^{-1}$ in laser-annealed (LA) PZT films on platinised silicon [20]. For comparison, this $d_{33,f}$ value was translated into $e_{33,f}$ (-3.9 C m$^{-2}$) by taking the typical value of Young's modulus (80 GPa) of PZT [55,56], which is slightly lower than those of flash lamp annealed films.

Given that the electrical properties are comparable, it is crucial to assess the processes. RTA necessitates a buffer and a processing time of tens of minutes and is incompatible with low-temperature glass. Laser annealing, on the other hand, requires non-transparent bottom electrodes (such as Pt or LaNiO$_3$) for nucleation, without which films cannot crystallize [19,20]. The small laser spot size also requires raster scanning of the laser beam, leading to inhomogeneities in the films. These limitations hinder process efficiency, resulting in reduced productivity. In contrast, the flash lamp process offers direct and rapid growth in an ambient environment without relying on any buffer or nucleation, while also being highly compatible with roll-to-roll production thanks to its large irradiation area. These advantages make it a high-output process.

The previous comparison and the results presented above provide evidence that the flash lamp process facilitates direct growth of piezoelectric films on a variety of glass substrates, which exhibit electromechanical properties meeting the requirements for piezoelectric applications, as demonstrated by the successful realisation of haptic rendering device. A distinctive advantage of the flash lamp process is its high compatibility with digital inkjet printing and large-scale roll-to-roll manufacturing, which puts it ahead of other low-temperature processes. Through the interaction of light with material, the process allows for growth of perovskite films without nucleation layers. Compared to aerosol-based deposition technique [57–60], the FLA-based process enables in-situ crystallization of perovskite phase and eliminates the need for post-annealing step to improve the properties.

As a conclusion, we proved in this paper that a well-defined Flash Lamp Annealing process enables the crystallization of piezoelectric perovskite thin films deposited on glass substrates that cannot withstand temperatures larger than 400°C. All these points are strong assets in favour of perovskite piezoelectric films for future transparent and flexible electronics.

## 4. Methods

### 4.1. Processing and deposition of PZT solution

$Pb(Zr_{0.53}Ti_{0.47})O_3$ solution was prepared via the standard 2-methoxyethanol (99.8 %, Sigma-Aldrich) route using freeze-dried lead(II) acetate trihydrate (99.99 %, Sigma-Aldrich), titanium(IV) isopropoxide (97 %, Sigma-Aldrich) and zirconium(IV) propoxide (70 % in propanol, Sigma-Aldrich) as starting compounds [61]. Concentration ($C_{Zr} + C_{Ti}$) was 0.3 M with a stoichiometry of metal cations Zr:Ti = 0.53:0.47, which corresponds to the morphotropic phase boundary composition with 10 mol.% Pb excess. Total volume of the solution was 50 mL. For more details see Ref. [61]. The solutions were spin coated on glass substrates for 30 s at 3000 rpm with an acceleration of 500 rpm s$^{-1}$. Three types of glasses were used: fused silica (2'' and 0.5 mm-thick, SIEGERT Wafer, fused silica) [62], AF32 (2'' and 0.3 mm-thick, SCHOTT, AF32) [49] and soda lime (38x26x1 mm$^3$, Marienfield, ref 1100020 (microscope slides) and 38x25x1 mm$^3$, Epredia, ref AB00000102E01MNZ10 (microscope slides)) [63]. Drying and pyrolysis were performed on hot plates for 2 min each at 130 °C and 350 °C, respectively. The deposition-drying-pyrolysis cycle was repeated four times to obtain 170 nm thick amorphous PZT films.

*4.2. Flash lamp annealing*

Crystallization of the amorphous films was performed with a flash lamp annealer (Pulseforge Invent, Novacentrix). The parameters to control the annealing process include energy density, pulse duration, and repetition rate. The respective values for a standard one-step process (fused silica and AF32 glasses) were 3 J cm$^{-2}$, 130 μs and 3.5 Hz. Note that the energy density was experimentally measured using a bolometer, and variation within 5 % in lamp energy density is observed in fast multipulse process due to the lamp stability. In a two-step process (soda lime glass), films were annealed for a few pulses at high power density followed by numerous pulses at lower power density. This two-step process is described in more details in Supplementary Note 4.1.

*4.3. Microstructural and optical characterization*

Phase composition and orientation of the films were investigated with GIXRD and standard $\theta$-$2\theta$ XRD, which were performed on a D8 Discover diffractometer (Bruker) using Cu-K$_\alpha$ radiation. The incidence angle of GIXRD was 0.5°. Both GIXRD and standard $\theta$-$2\theta$ XRD patterns were recorded in the $2\theta$-range from 20° to 60° with a step-size of 0.02°.

Samples for STEM analyses were prepared by grinding, dimpling and final Ar milling (Gatan PIPS Model 691, New York, NY, USA). STEM studies were carried out using a probe Cs-corrected Jeol ARM200 CF (Jeol, Tokyo, Japan) equipped with Centurio energy-dispersive X-ray spectroscopy system (Jeol, Tokyo, Japan), operated at 200kV.

Optical spectra were obtained using a UV/Vis spectrophotometer (Spectro L1050, PerkinElmer).

*4.4. Electromechanical characterization*

Interdigitated Pt electrodes were deposited and patterned on top of the films with lift-off photolithography using direct laser writing (MLA 150, Heidelberg Instruments), as described in our previous work [42]. Pt was chosen as it is the most mature electrode material for piezoelectric thin films. Two sizes of IDE electrodes were used. Finger width, length of digits facing each other, interdigital gap, and pairs of digits were 5 μm (5 μm),1730 μm (370 μm), 3 μm (3 μm), and 615 (50), respectively for large (small) IDEs100 nm-thick Pt electrodes were sputtered with a MED 020 metallizer (BalTec).

Electrical characterization was performed using a TF Analyzer 2000 (aixACCT). The polarization was measured as a function of electric field $P(\mathbf{E})$ with a triangular waveform at 100 or 10 Hz. Bipolar fatigue cycling was performed at 1 kHz, with a triangular fatigue voltage of 100 V.

The converse piezoelectric response of 500 nm PZT film on fused silica glass (Figure 3b) was measured with a thin-film sample holder unit (aixACCT) and an interferometer (SP120/2000, SIOS) [64]. Cantilevers with dimensions of 25 × 3.4 mm$^2$ were cut from the wafers using a wire saw. Geometry of the IDE capacitor was the same as for large IDE. The converse piezoelectric coefficient $e_{33,f}$ was extracted as described in Ref [65]. A Young's modulus of 73 GPa has been used for the fused silica substrate.

Relative permittivity $\varepsilon_r$ and dielectric losses tan$\delta$ of 1 μm-thick film were measured as functions of DC voltage with a probing AC signal of 0.5-1 V at 1 kHz. The IDE used for this measurement (Figure 3c) has finger width of 10 μm, length of digits facing each other of 872 μm, interdigital gap of 10 μm, and 502 pairs of digits, respectively, corresponding to an effective area of 0.86 mm$^2$. Capacitance $C$ and tan$\delta$ were measured as functions of frequency $f$ with an Agilent E4990A impedance analyzer with an AC voltage of 1 V. Corresponding $\varepsilon_r$ was calculated from capacitance. The IDE used for this measurement (Figure 3d) has the respective parameters of 10 μm, 2340 μm, 10 μm and 51, respectively, corresponding to an effective area of 0.209 mm$^2$.

*4.5. Haptic device*

The 1-μm-thick PZT film was prepared on AF32 glass, as described above. Based upon FEM, the IDE Pt electrodes had the overall dimension of 2050 × 2340 μm$^2$ and were distributed on the 15.4 × 3 mm$^2$ device as schematically shown in Figure 4a. Copper wires bonded using silver epoxy were used to connect voltage and ground electrode pads. The haptic device was placed on suspended flexible foam tape and was connected to a waveform generator (33210A,

Keysight) via an amplifier (WMA-300, Falco Systems). The out-of-plane displacement was recorded using a laser doppler vibrometer (OFV-5000, Polytec). A computer-controlled *x-y* stage was used to move the haptic device for line scans and 2D mapping. The whole set-up was controlled with the vibrometer via a LabVIEW program, which have been also demonstrated in a previous paper [42]. Note that the two actuators were connected in parallel, therefore, the voltage applied to each actuator is the same. Note also that all the measurements in Figure 4 were done with a unipolar voltage.

**Data Availability**

The data that supports the findings of the study are included in the main text and supplementary information files. The raw data generated in this study have been deposited in the Zenodo repository [66] under DOI 10.5281/zenodo.10622727.

**References**

1.   Muralt, P. Recent progress in materials issues for piezoelectric MEMS. *J. Am. Ceram. Soc.* **91**, 1385–1396 (2008).

2.   Defaÿ, E. *Integration of ferroelectric and piezoelectric thin films*. (2011). doi:10.1002/9781118616635.

3.   Gao, X. *et al.* Giant piezoelectric coefficients in relaxor piezoelectric ceramic PNN-PZT for vibration energy harvesting. *Adv. Funct. Mater.* **28**, 1706895 (2018).

4.   Yeo, H. G. *et al.* Strongly (001) oriented bimorph PZT film on metal foils grown by rf-sputtering for wrist-worn piezoelectric energy harvesters. *Adv. Funct. Mater.* **28**, 1–9 (2018).

5.   Glinsek, S. *et al.* Inkjet-printed piezoelectric thin films for transparent haptics. *Adv. Mater. Technol.* **7**, 2200147 (2022).

6.   Mertin, S. *et al.* Piezoelectric and structural properties of c-axis textured aluminium scandium nitride thin films up to high scandium content. *Surf. Coatings Technol.* **343**, 2–6 (2018).

7.   Godard, N. *et al.* 1-mW vibration energy harvester based on a cantilever with printed polymer multilayers. *Cell Reports Phys. Sci.* **1**, 100068 (2020).

8.   Dubois, M. A., Muralt, P. & Plessky, V. BAW resonators based on aluminum nitride thin films. *Proc. IEEE Ultrason. Symp.* **2**, 907–910 (1999).

9.   Bassiri-Gharb, N., Bastani, Y. & Bernal, A. Chemical solution growth of ferroelectric oxide thin films and nanostructures. *Chem. Soc. Rev.* **43**, 2125–2140 (2014).

10.  Brennecka, G. L., Ihlefeld, J. F., Maria, J. P., Tuttle, B. A. & Clem, P. G. Processing

technologies for high-permittivity thin films in capacitor applications. *J. Am. Ceram. Soc.* **93**, 3935–3954 (2010).

11. Yu, M. J. *et al.* Amorphous InGaZnO thin-film transistors compatible with roll-to-roll fabrication at room temperature. *IEEE Electron Device Lett.* **33**, 47–49 (2012).

12. Song, L., Glinsek, S. & Defay, E. Toward low-temperature processing of lead zirconate titanate thin films: Advances, strategies, and applications. *Appl. Phys. Rev.* **8**, 041315 (2021).

13. Park, J. W., Kang, B. H. & Kim, H. J. A review of low-temperature solution-processed metal oxide thin-film transistors for flexible electronics. *Adv. Funct. Mater.* **30**, 1904632 (2020).

14. Joe, D. J. *et al.* Laser-material interactions for flexible applications. *Adv. Mater.* **29**, 1606586 (2017).

15. Bretos, I. *et al.* Activated solutions enabling low-temperature processing of functional ferroelectric oxides for flexible electronics. *Adv. Mater.* **26**, 1405–1409 (2014).

16. Bretos, I. *et al.* Low-temperature crystallization of solution-derived metal oxide thin films assisted by chemical processes. *Chem. Soc. Rev.* **47**, 291–308 (2018).

17. Zhang, X. D. *et al.* Low-temperature preparation of sputter-deposited Pb(Zr$_{0.52}$Ti$_{0.48}$)O$_3$ thin films through high oxygen-pressure annealing. *J. Cryst. Growth* **310**, 783–787 (2008).

18. Tue, P. T., Shimoda, T. & Takamura, Y. A facile solution-combustion-synthetic approach enabling low-temperature PZT thin-films. *APL Mater.* **8**, 21112 (2020).

19. Kang, M. G. *et al.* Direct growth of ferroelectric oxide thin films on polymers through laser-induced low-temperature liquid-phase crystallization. *Chem. Mater.* **32**, 6483–

6493 (2020).

20. Fink, S., Lübben, J., Schneller, T., Vedder, C. & Böttger, U. Impact of the processing temperature on the laser-based crystallization of chemical solution deposited lead zirconate titanate thin films on short timescales. *J. Appl. Phys.* **131**, 125302 (2022).

21. Prucnal, S., Rebohle, L. & Skorupa, W. Doping by flash lamp annealing. *Mater. Sci. Semicond. Process.* **62**, 115–127 (2017).

22. Schatz, A., Pantel, D. & Hanemann, T. Towards low-temperature deposition of piezoelectric Pb(Zr,Ti)O$_3$: Influence of pressure and temperature on the properties of pulsed laser deposited Pb(Zr,Ti)O$_3$. *Thin Solid Films* **636**, 680–687 (2017).

23. Greer, J. Large-area commercial pulsed laser deposition. in *Pulsed laser deposition of thin films* (ed. Eason, R.) 191–213 (John Wiley \& Sons, Inc., 2007). doi:10.1002/9780470052129.ch9.

24. Pavageau, F. *et al.* Highly transparent piezoelectric PZT membranes for transducer applications. *Sensors Actuators A Phys.* **346**, 113866 (2022).

25. Angmo, D., Larsen-Olsen, T. T., Jørgensen, M., Søndergaard, R. R. & Krebs, F. C. Roll-to roll inkjet printing and photonic sintering of electrodes for ITO free polymer solar cell modules and facile product integration. *Adv. Energy Mater.* **3**, 172–175 (2013).

26. Rebohle, L., Prucnal, S. & Reichel, D. *Flash lamp Annealing: From basics to applications*. (Springer International Publishing, 2019). doi:10.1007/978-3-030-23299-3.

27. Gilshtein, E. *et al.* Inkjet-printed conductive ITO patterns for transparent security systems. *Adv. Mater. Technol.* **5**, 2000369 (2020).

28. Shin, J. H. *et al.* A flash-induced robust Cu electrode on glass substrates and its

application for thin-film μLEDs. *Adv. Mater.* **33**, 2007186 (2021).

29. Loganathan, K. *et al.* Rapid and up-scalable manufacturing of gigahertz nanogap diodes. *Nat. Commun. 2022* **13**, 3260 (2022).

30. Weidling, A. M., Turkani, V. S., Luo, B., Schroder, K. A. & Swisher, S. L. Photonic curing of solution-processed oxide semiconductors with efficient gate absorbers and minimal substrate heating for high-performance thin-film transistors. *ACS Omega* **6**, 17323–17334 (2021).

31. Yarali, E. *et al.* Low-voltage heterojunction metal oxide transistors via rapid photonic processing. *Adv. Electron. Mater.* **6**, 2000028 (2020).

32. Daunis, T. B., Schroder, K. A. & Hsu, J. W. P. P. Photonic curing of solution-deposited ZrO2 dielectric on PEN: a path towards high-throughput processing of oxide electronics. *Flex. Electron.* **4**, 7 (2020).

33. Sánchez, S., Jerónimo-Rendon, J., Saliba, M. & Hagfeldt, A. Highly efficient and rapid manufactured perovskite solar cells via Flash InfraRed Annealing. *Mater. Today* **35**, 9–15 (2020).

34. Jung, D. H. *et al.* Flash-induced ultrafast recrystallization of perovskite for flexible light-emitting diodes. *Nano Energy* **61**, 236–244 (2019).

35. Yao, Y. *et al.* Direct processing of $PbZr_{0.53}Ti_{0.47}O_3$ films on glass and polymeric substrates. *J. Eur. Ceram. Soc.* **40**, 5369–5375 (2020).

36. Basdogan, C., Giraud, F., Levesque, V. & Choi, S. A review of surface haptics: enabling tactile effects on touch surfaces. *IEEE Trans. Haptics* **13**, 450–470 (2020).

37. Bernard, F., Gorisse, M., Casset, F., Chappaz, C. & Basrour, S. Design, fabrication and characterization of a tactile display based on aln transducers. *Procedia Eng.* **87**, 1310–

1313 (2014).

38. Wiertlewski, M., Friesen, R. F. & Colgate, J. E. Partial squeeze film levitation modulates fingertip friction. *Proc. Natl. Acad. Sci. U. S. A.* **113**, 9210–9215 (2016).

39. Bernard, F. Conception, fabrication et caractérisation d'une dalle haptique à base de microactionneurs piézoélectriques. (Université Grenoble Alpes, 2016).

40. Sednaoui, T. *et al.* Experimental Evaluation of Friction Reduction in Ultrasonic Devices. *IEEE World Haptics Conf. WHC 2015* 37–42 (2015) doi:10.1109/WHC.2015.7177688.

41. Kwok, C. K. & Desu, S. B. Pyrochlore to perovskite phase transformation in sol-gel derived lead-zirconate-titanate thin films. *Appl. Phys. Lett.* **60**, 1430–1432 (1992).

42. Glinsek, S. *et al.* Fully transparent friction-modulation haptic device based on piezoelectric thin Film. *Adv. Funct. Mater.* **30**, 2003539 (2020).

43. ICDD database PDF4+ v.19. (2019).

44. Gueye, I. *et al.* Chemistry of surface nanostructures in lead precursor-rich $PbZr_{0.52}Ti_{0.48}O_3$ sol–gel films. *Appl. Surf. Sci.* **363**, 21–28 (2016).

45. Fengler, F. P. G. *et al.* Domain pinning: Comparison of hafnia and PZT based ferroelectrics. *Adv. Electron. Mater.* **3**, 1600505 (2017).

46. Nguyen, C. H. *et al.* Probing-models for interdigitated electrode systems with ferroelectric thin films. *J. Phys. D. Appl. Phys.* **51**, 175303 (2018).

47. Damjanovic, D. Ferroelectric, dielectric and piezoelectric properties of ferroelectric thin films and ceramics. *Reports Prog. Phys.* **61**, 1267–1324 (1998).

48. Casset, F. *et al.* Low voltage haptic slider built using sol-gel thin-film PZT actuators reported on glass. in *2019 IEEE 32nd International Conference on Micro Electro*

*Mechanical Systems (MEMS)* 990–993 (IEEE, 2019). doi:10.1109/MEMSYS.2019.8870715.

49. AF32 glass datasheet. https://www.pgo-online.com/intl/af32.html (2022).

50. Song, L. *et al.* Piezoelectric thick film for power-efficient haptic actuator. *Appl. Phys. Lett.* **121**, 212901 (2022).

51. Yamakawa, K. *et al.* Novel Pb(Ti, Zr)O3 (PZT) crystallization technique using flash lamp for ferroelectric RAM (FeRAM) embedded LSIs and one transistor type FeRAM devices. *Japanese J. Appl. Physics, Part 1 Regul. Pap. Short Notes Rev. Pap.* **41**, 2630–2634 (2002).

52. Palneedi, H. *et al.* Intense pulsed light thermal treatment of Pb(Zr,Ti)O$_3$/metglas heterostructured films resulting in extreme magnetoelectric coupling of over 20 V cm$^{-1}$ Oe$^{-1}$. *Adv. Mater.* 2303553 (2023) doi:10.1002/adma.202303553.

53. Ouyang, J., Cormier, D., Williams, S. A. & Borkholder, D. A. Photonic sintering of aerosol jet printed lead zirconate titanate (PZT) thick films. *J. Am. Ceram. Soc.* **99**, 2569–2577 (2016).

54. Ledermann, N. *et al.* {100}-textured, piezoelectric Pb(Zr$_x$, Ti$_{1-x}$)O$_3$ thin films for MEMS: integration, deposition and properties. *Sensors Actuators A Phys.* **105**, 162–170 (2003).

55. Casset, F. *et al.* Young modulus and Poisson ratio of PZT thin film by picosecond ultrasonics. in *IEEE International Ultrasonics Symposium, IUS* 2180–2183 (2012). doi:10.1109/ULTSYM.2012.0544.

56. Cohen, R. E., Heifets, E., Fu, H. & Phys Lett, A. First-principles computation of elasticity of Pb(Zr, Ti)O3: The importance of elasticity in piezoelectrics. *AIP Conf. Proc.*

**582**, 11–22 (2001).

57. Sadl, M. et al. Energy-storage-efficient 0.9Pb(Mg$_{1/3}$Nb$_{2/3}$)O$_3$–0.1PbTiO$_3$ thick films integrated directly onto stainless steel. *Acta Mater.* **221**, 117403 (2021).

58. Patil, D. R. et al. Piezoelectric thick film deposition via powder/granule spray in vacuum: A review. *Actuators* **9**, 59 (2020).

59. Hwang, G. T. et al. Self-powered wireless sensor node enabled by an aerosol-deposited PZT flexible energy harvester. *Adv. Energy Mater.* **6**, 1–9 (2016).

60. Sadl, M. et al. Multifunctional energy storage and piezoelectric properties of 0.65Pb(Mg$_{1/3}$Nb$_{2/3}$)O$_3$-0.35PbTiO$_3$ thick films on stainless-steel substrates. *J. Phys. Energy* **4**, 0–9 (2022).

61. Godard, N., Glinsek, S. & Defay, E. Inkjet-printed silver as alternative top electrode for lead zirconate titanate thin films. *J. Alloys Compd.* **783**, 801–805 (2019).

62. Corning. Corning® HPFS® 7979, 7980, 8655 Fused Silica, Optical Materials Product Information. https://www.corning.com/media/worldwide/csm/documents/HPFS_Product_Brochure_All_Grades_2015_07_21.pdf (2015).

63. See more information about soda lime glass used for common microscope slides: https://www.marienfeld-superior.com/microscopes-slides-special-size-76-x-52-mm.html (2022).

64. Mazzalai, A., Balma, D., Chidambaram, N., Matloub, R. & Muralt, P. Characterization and fatigue of the converse piezoelectric effect in PZT films for MEMS applications. *J. Microelectromechanical Syst.* **24**, 831–838 (2015).

65. Nguyen, C. H. et al. Probing-models for interdigitated electrode systems with

ferroelectric thin films. *J. Phys. D. Appl. Phys.* **51**, (2018).

66. Song, L. *et al.* Raw data for Crystallization of piezoceramic films on glass via glash lamp annealing. DOI: 10.5281/zenodo.10622727 (2024).

67. Song, L. Processing of piezoelectric oxide films for surface haptics. (Université du Luxembourg, 2023).

**Acknowledgements**

Luxembourg National Research Fund (FNR) is acknowledged for the financial support through project FLASHPOX (C21/MS/16215707, grant recipient: S.Gli.). L.S. and A.B.M. acknowledge financial support from the FNR under the project PACE (PRIDE/17/12246511/PACE). A.B. an B.K. acknowledge funding by the Slovenian Research Agency (projects J2-3041, J2-2497, grant recipient: A.B.). Barnik Mandal and Poorani Gnanasambandan are acknowledged for performing SEM. Some of the data from this manuscript was previously published in the thesis of L.S.[67].

**Author Contributions**

L.S. and J.C. contributed equally to the acquisition, analysis, interpretation of data and writing of the work. A.B. contributed to the acquisition and analysis of data. A.B.M., B.K. and S.G. contributed to the acquisition of data. E.D. contributed to the conception of the work. S.Gli. contributed to the conception of the work and substantively revised it.

**Competing Interests**

The authors declare no competing interests.

Supplementary Information for

# Crystallization of piezoceramic films on glass via flash lamp annealing

Longfei Song,[1,2,†] Juliette Cardoletti,[1,†] Alfredo Blázquez Martínez,[1,2] Andreja Benčan,[3] Brigita Kmet,[3] Stéphanie Girod,[1] Emmanuel Defay,[1] Sebastjan Glinšek[1,*]

[1] Materials Research and Technology Department, Luxembourg Institute of Science and Technology, 41 rue du Brill, L-4422 Belvaux, Luxembourg

[2] University of Luxembourg, 41 rue du Brill, L-4422 Belvaux, Luxembourg

[3] Electronic Ceramics Department, Jožef Stefan Institute, Jamova cesta 39, 1000 Ljubljana, Slovenia

[†] These authors contributed equally: L. S. and J. C.

*Corresponding author: sebastjan.glinsek@list.lu

**Contents**

## Supplementary Note 1: Finite element modelling (FEM)

In order to investigate the feasibility of crystallization of $PbZr_{0.53}Ti_{0.47}O_3$ (PZT) thin films via flash lamp annealing, finite element modelling was performed using surface mode of SimPulse software to study the temperature evolution across the thickness of the samples[1].

To estimate the absorbance inside amorphous PZT films, a method using a bolometer was utilized[2]. While this method is not ideal, it is a reasonable approach as the measurement is performed under the same conditions as flash lamp annealing. First, the energy density on the surface of the stage of the flash-lamp annealer was measured with a bolometer after the light passed through air (direct exposure to light, $E_{tot}$ = 2.8 J cm$^{-2}$), bare fused silica glass (estimation of light reflected from the glass substrate assuming negligible absorption, $E_{glass} = E_{tot} - E_{ref}$ = 2.6 J cm$^{-2}$), and a fused silica glass with amorphous PZT on top (estimation of light that passes through the stack, $E_{sample} = E_{trans}$ = 1.8 J cm$^{-2}$), see Supplementary Figure 1. To estimate the quantity of absorbed light, we used the equation:

$$E_{abs} = E_{tot} - E_{sample} - E_{ref} = E_{tot} - E_{trans} - (E_{tot} - E_{glass}). \qquad (1)$$

In equation (1), $E$ denotes energy density in J cm$^{-2}$. The resulting absorbed energy $E_{abs}$ is 0.8 J cm$^{-2}$. The corresponding absorption value $A$ is estimated as:

$$A = \frac{E_{abs}}{E_{tot}}, \qquad (2)$$

which was calculated as 28.5 %. Note that this value is a rough estimation only.

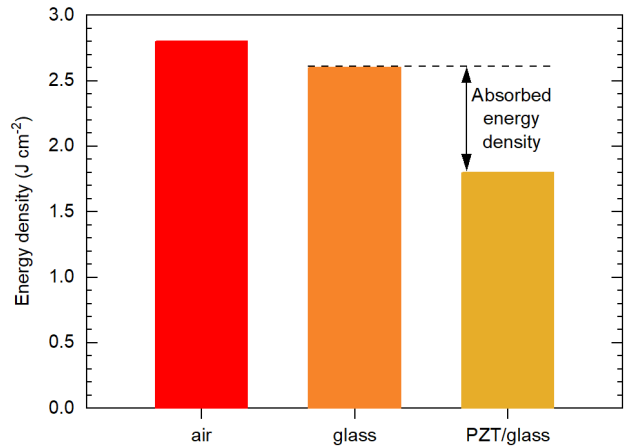

Supplementary Figure 1. **Transmitted energy density** measured with a bolometer after the light passing through: air (empty chamber), 500 µm-thick fused silica glass, and pyrolyzed PZT/fused silica stack. The energy density and the length of the applied light pulse were 2.8 J cm$^{-2}$ and 130 µs.

## Supplementary Note 2: Films grown on fused silica glass

### Supplementary Note 2.1: Phase composition and microstructural characterizations

A standard $\theta$-$2\theta$ X-ray diffraction (XRD) pattern of a flash lamp annealed 1 µm-thick PZT film on fused silica is shown in Supplementary Figure 2. Presence of piezoelectrically active perovskite phase is revealed, and all the reflections can be identified with the PZT powder diffraction file (PDF) No 01-070-4264[3]. Reflections of secondary phases are not observed.

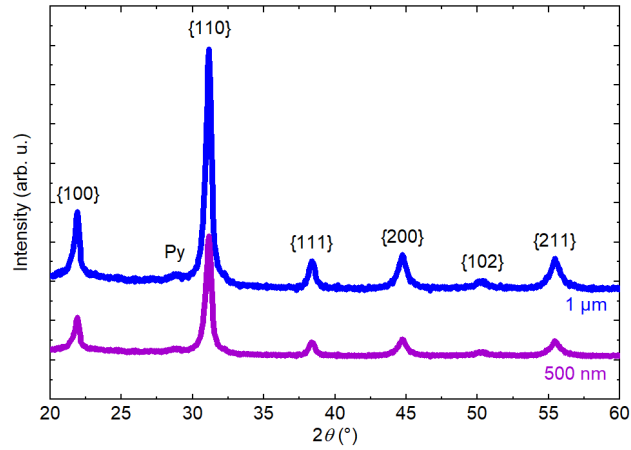

Supplementary Figure 2. **XRD study**. θ-2θ XRD patterns of flash lamp annealed 500 nm and 1 μm PZT films on fused silica glass. The films were processed with 50 pulses per layer (3 and 6 crystallizations in total, respectively). Parameters of each pulse are reported in caption of Fig. 2 in the main manuscript. The PDFs No 01-070-4264 and 04-014-5162[3] have been used to identify the perovskite and pyrochlore phases, respectively.

Grazing incidence X-ray diffraction (GIXRD) pattern of the film annealed in a conventional RTA furnace at 700 °C is shown in Supplementary Figure 3. Only pyrochlore phase is detected and cracks appeared on the surface, as previously reported in Ref.[4].

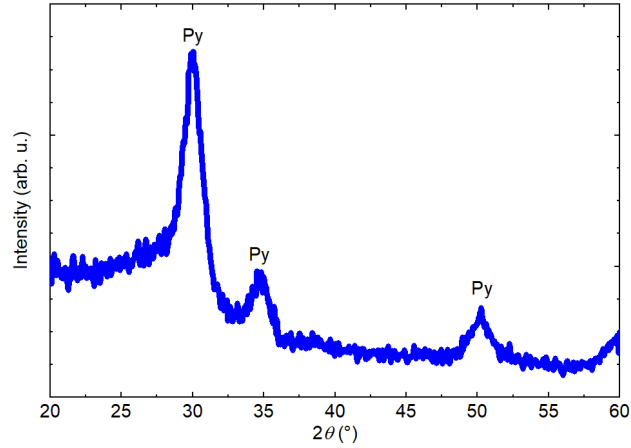

Supplementary Figure 3. **XRD study**. GIXRD pattern of a conventionally RTA-processed film on fused silica at 700 °C. Py indicates the pyrochlore reflection according to PDF No 04-014-5162[3].

The transmittance spectra of PZT thin films of various thicknesses on fused silica glass are shown in Supplementary Figure 4. The inset shows that 1 µm-thick PZT films are transparent.

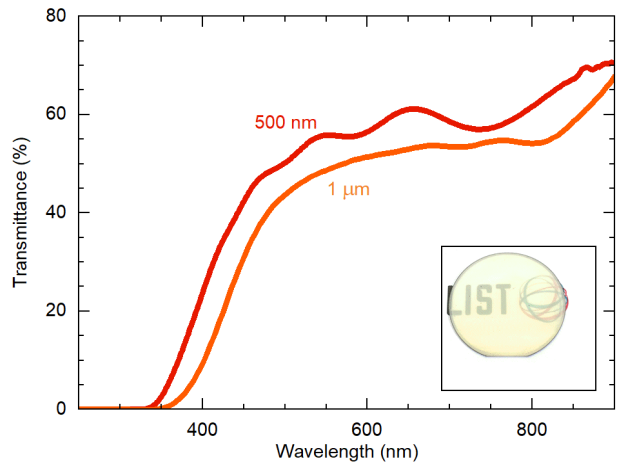

Supplementary Figure 4. **Transmittance of 500 nm and 1 µm-thick PZT thin films on fused silica glass.** Inset shows the optical appearance of the 1 µm-thick film. The films were processed with 50 pulses per layer (3 and 6 crystallizations in total, respectively). Parameters of each pulse are reported in caption of Fig. 2 in the main manuscript.

Supplementary Figure 5 displays cross-sectional scanning electron microscopy (SEM) image of a flash lamp annealed 1 µm-thick PZT, showing a clean interface between PZT film and glass. The microstructure is granular with present porosity.

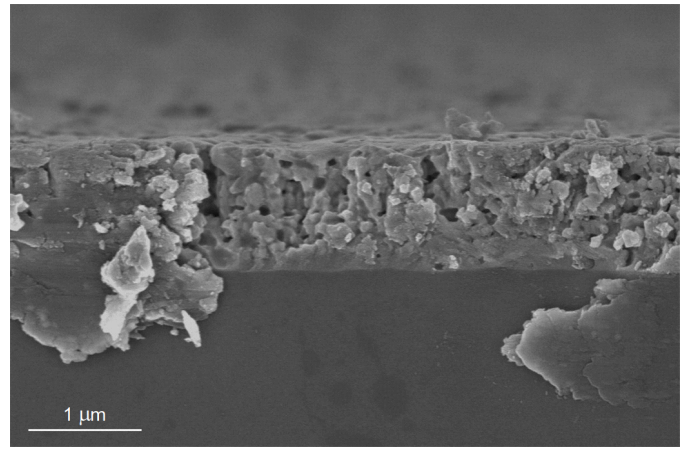

Supplementary Figure 5. **Microstructure characterization**. Cross-sectional SEM image of flash lamp annealed 1 µm PZT film on fused silica glass. Note that the large flake-like structures observed in the glass are due to glass cleaving.

A detailed TEM analysis of the 170 nm-thick PZT thin film on fused silica glass is shown in Supplementary Figure 6.

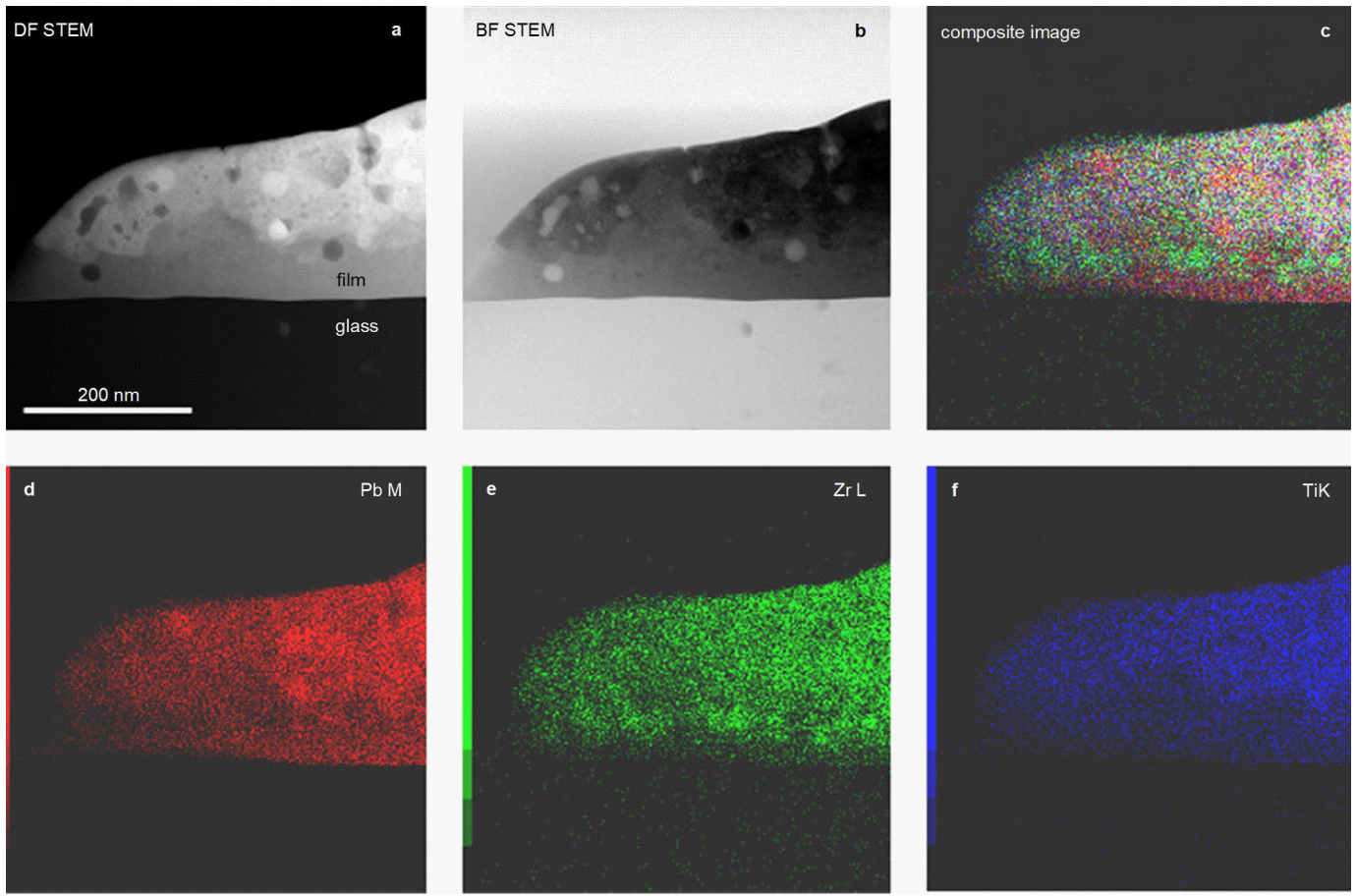

Supplementary Figure 6. **TEM analysis of 170 nm PZT thin film on fused silica glass**. a) Cross sectional dark-field (DF) and b) bright-field (BF) scanning transmission electron microscope (STEM) image with c-f) corresponding energy-dispersive X-ray spectroscopy system mapping showing porous, chemically non-homogenous film. Pores are darker/brighter spots on DF/BF STEM images, correspondingly.

**Supplementary Note 2.2: Electrical measurements**

*Supplementary Note 2.2.1: 170 nm-thick film grown on fused silica glass*

We found that the films annealed with 50 pulses exhibit the optimal electrical properties. Polarization as a function of electric field $P(\mathbf{E})$ and corresponding current density $J(\mathbf{E})$ loops of the 170 nm-thick films treated with 50 pulses at 100 and 10 Hz are shown in Supplementary Figure 7. The polarization loops of the films are initially pinched, before opening up during (wake-up effect) upon electric-field cycling (1.1×10⁶ bipolar cycles). At 10 Hz, the $P_r$ is 10 μC cm⁻², and two sharp peaks in $J(E)$, linked to ferroelectric switching, are observed.

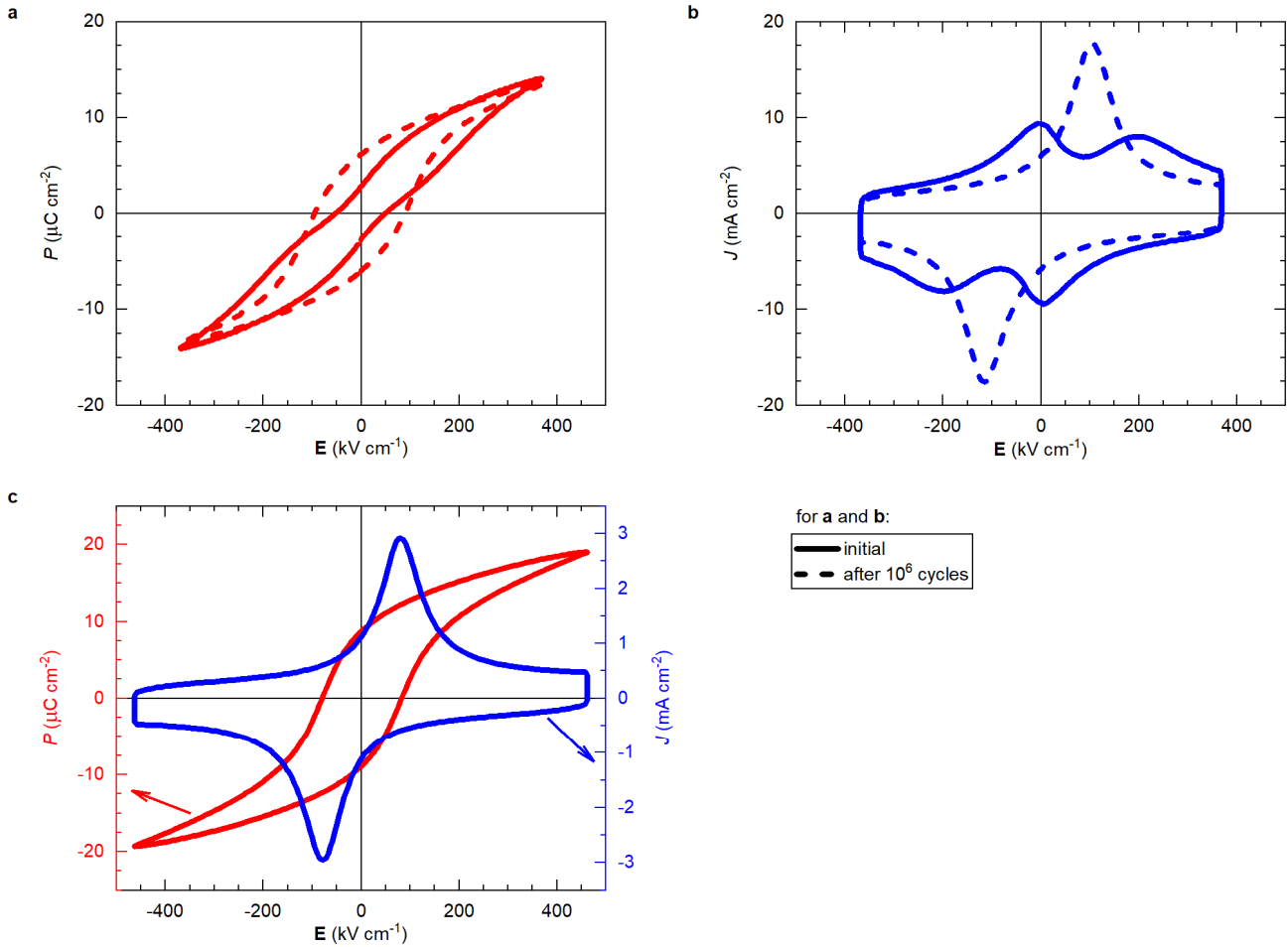

Supplementary Figure 7. **Ferroelectric characterization**. a) Polarization $P(E)$ hysteresis loops of the 50 pulses annealed 170 nm-thick PZT film on fused silica before and after 1.1×10⁶ wake-up cycles, measured at 100 Hz. b) corresponding current density loops $J(E)$. c) $P(E)$ and $J(E)$ measured at 10 Hz, by taking the average after 500 cycles. Large interdigitated electrodes (IDE) were used in these measurements, corresponding to an effective area of 0.36 mm².

*Supplementary Note 2.2.2: 500 nm-thick film grown on fused silica glass*

The $P(E)$ and $J(E)$ loops of the 500 nm-thick PZT film on fused silica glass are shown in Supplementary Figure 8a with a maximum polarization $P_{max}$ of 21 µC cm$^{-2}$ and a remanent polarization $P_r$ of 11 µC cm$^{-2}$. Its coercive field $E_c$ is 95 kV cm$^{-1}$. Note that these values were obtained after $10^3$ wake-up cycles. The displacement of a cantilever structure shows a typical butterfly loop (Supplementary Figure 8b). At 150 V the vertical displacement at the free end of the cantilever is 625 nm, corresponding to a piezoelectric coefficient $e_{33,f}$ of -5 C m$^{-2}$ [5].

All the results confirm good ferroelectric properties of 50 pulses flash lamp annealed films.

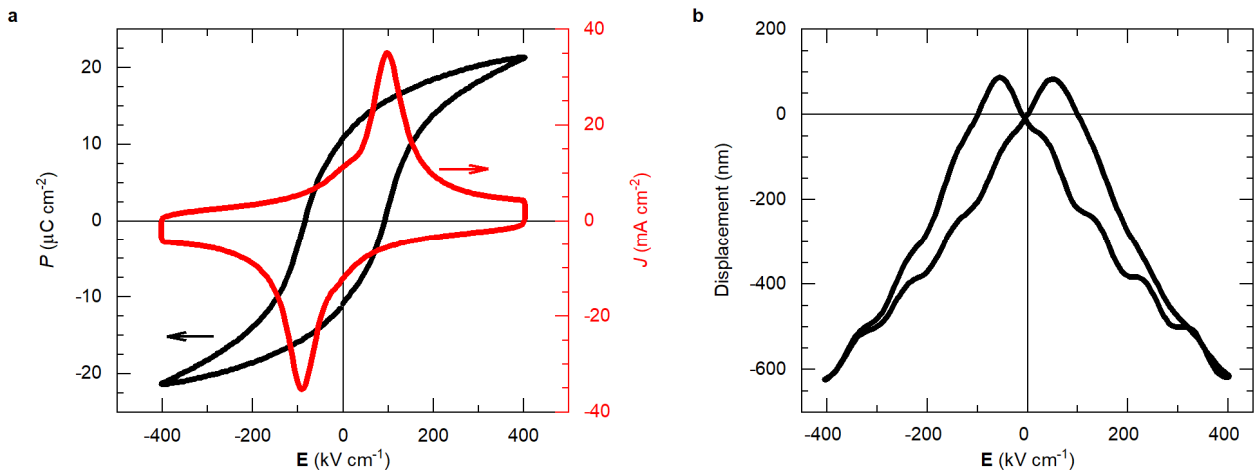

Supplementary Figure 8. **Electromechanical characterization of a 500 nm-thick PZT film on fused silica glass**. a) Ferroelectric and b) displacement characterizations of the PZT film at 100 Hz and 11 Hz, respectively. The samples were processed with 50 pulses per layer with energy density, pulse duration and repetition rate of 3 J cm$^2$, 130 µs, and 3.5 Hz, respectively.

**Supplementary Note 2.3: Comparison of properties at different thicknesses**

To have better overview of the results, a table with ferroelectric, piezoelectric and optical properties of the 170 nm, 500 nm and 1 µm-thick films on fused silica substrates is provided in Supplementary Table 1.

Supplementary Table 1. **Properties of 170 nm, 500 nm and 1 µm-thick PZT films on fused silica glass**. Remanent and maximum polarization ($P_r$ and $P_{max}$) at an applied voltage of 150 V, relative permittivity and dielectric losses ($\varepsilon_r$ and $\tan\delta$), piezoelectric coefficient $e_{33,f}$, and transmittance ($T$) at a wavelength of 550 nm. The films were processed with 50 pulses per layer with energy density, pulse duration and repetition rate of 3 J cm$^{-2}$, 130 µs, and 3.5 Hz, respectively. Large IDEs were used in these measurements, corresponding to an effective area of 0.36 mm$^2$, 1.08 mm$^2$, and 2.16 mm$^2$ for 170 nm, 500 nm and 1 µm, respectively.

| Film thickness | $P_r$ (µC cm$^{-2}$) | $P_{max}$ (µC cm$^{-2}$) | $\varepsilon_r$ | $\tan\delta$ | $e_{33,f}$ (C m$^{-2}$) | $T$ (%) |
|---|---|---|---|---|---|---|
| 170 nm | 10 | 19 | 200 | 0.05 | -2 | 64 |
| 500 nm | 11 | 21 | 270 | 0.05 | -5 | 56 |
| 1 µm | 12 | 24 | 450 | 0.05 | -5 | 49 |

**Supplementary Note 3: Thick PZT film on AF32 glass for surface haptic device**

**Supplementary Note 3.1: Phase composition and microstructural characterization**

A 1 µm-thick PZT film was grown on AF32 glass. Process parameters were the same as for the films on fused silica, namely 3 J cm$^{-2}$ in energy density, 130 µs pulse duration and 50 pulses with a repetition rate of 3.5 Hz. Supplementary Figure 9 shows $\theta$-$2\theta$ XRD pattern of the film. Reflections of the perovskite phase are revealed and all of them can be identified with the PZT PDF No 01-070-4264[3], as for the film grown on fused silica. Note that the reflection at around 39° comes from the Pt electrodes.

Supplementary Figure 10 shows cross-sectional SEM image of a 1 µm-thick PZT film on AF32 glass, used for haptic device. It reveals a dense and granular microstructure, and also a clear interface between the film and glass.

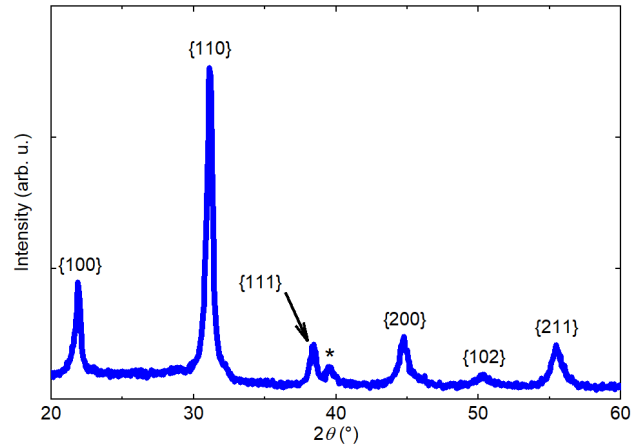

Supplementary Figure 9. **XRD study**. $\theta$-$2\theta$ XRD pattern of flash lamp annealed 1 µm thick PZT film on AF32 glass used for the haptic device. * denotes the signal of Pt IDEs on top of the PZT film. The PDFs No 01-070-4264[3] has been used to identify the perovskite phase.

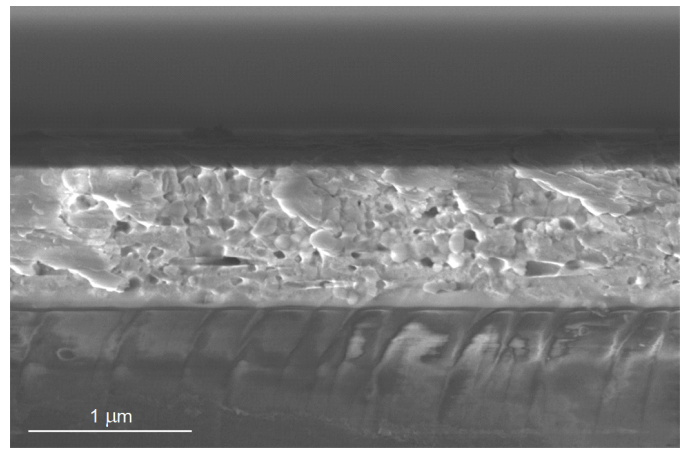

Supplementary Figure 10. **Microstructure characterization**. Cross-sectional SEM image of 1 µm-thick PZT on AF32 glass, used for haptic device. The processing conditions have been described in the Methods.

**Supplementary Note 3.2: Haptic devices**

*Supplementary Note 3.2.1: Thickness of piezoelectric film*

In general, piezoelectric films show increased electromechanical response with increasing thickness due to larger contributions of domains[6]. In the case of piezoelectric actuators with interdigitated (IDE) geometry, additional benefit of using thicker film is an increased in-plane force $F_3$ exerted by the piezoelectric layer upon applied electric field $E_3$. The force is expressed as:

$$F_3 = -e_{33,f} E_3 A, \qquad (2)$$

where $E_3$ is an in-plane electric field, $e_{33,f}$ is an effective piezoelectric coefficient and $A$ is a cross-section. In IDE geometry $E_3$ roughly equals to an applied voltage $U$ divided by a gap $a$ between the fingers, while $A$ equals to a film thickness $t_f$ multiplied by a finger length $l$. The above equation can be therefore re-written as:

$$F_3 = -e_{33,f} \frac{U}{a} l t_f, \qquad (3)$$

from which it follows that the force exerted by IDE piezoelectric actuator (at constant voltage) can be increased by increasing film's thickness (and decreasing the gap between the fingers). Considering these points and ease of processing, we defined 1 μm-thick PZT film as a good compromise.

*Supplementary Note 3.2.2: Finite element modelling*

Two-dimension (2D) FEM was carried out using COMSOL software to design a haptic device. 1 μm PZT/AF32 glass structure was used in the modelling with the total length of 15.4 mm. Two symmetric actuating areas were created with IDEs with 129 pairs of digits and a spacing of 8.4 mm. The width of the fingers and the interdigital gap are 5 μm and 3 μm, respectively.

Young's modulus and Poisson ratio for AF32 glass are 74.8 GPa and 0.238, respectively[7]. Influence of the electrodes on the deflection were ignored due to their lower thicknesses. The effective transverse piezoelectric coefficient $e_{33,\text{eff}}$ and relative permittivity $\varepsilon_r$ were set to

-4.5 C m$^{-2}$ and 400, respectively. Note that $e_{33,\text{eff}}$ was extracted from the modelling by matching the experimental displacement value of the haptic device, after having measured the damping loss factor $\eta^8$, which was obtained by sweeping the frequency of the actuator and collecting the displacement, as shown in Fig. 4b. $\eta$ corresponds to the breadth of the resonance peak ($\Delta f/f$), where $\Delta f$ is the full-width-at-half-maximum and $f$ is the resonance frequency. The obtained value of $\eta$ is 0.0163.

At a simulated resonant frequency of 40.2 kHz, the device exhibits a maximum displacement (1.7 µm peak-to-peak) when driven with 60 V. Supplementary Figure 11 illustrates the device's mode shape at resonance, specifically a Lamb wave mode featuring four equally spaced nodes along its length.

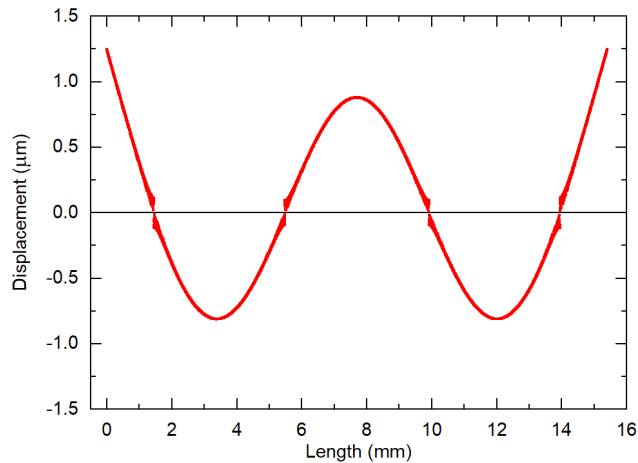

Supplementary Figure 11. **Finite element modelling of haptic device**. Wave shape along the length of the device at resonant frequency.

*Supplementary Note 3.2.3: Device performance*

The device is operating in a resonance mode and the standing wave correspond to the anti-symmetric ($A_0$) Lamb wave, which has been considered as optimal for piezoelectric haptics in previous works[9]. Lamb waves in glass plates for haptics were extensively studied by Bernard[10]. Figure II.3 on page 47 from his work[10] shows wavevectors of Lamb waves as functions of a frequency-thickness product. In our work the glass is 300 µm-thick and is operated at a frequency of 40.2 kHz, leading to the frequency-thickness product of 0.012 MHz mm. This is below the appearance of any other modes than $A_0$. The fact that Bernard used EAGLE XG and we are using AF32 glass does not change the outcome of this analysis as both glasses have comparable mechanical properties and density.

Performance of the device is compared to other piezoelectric haptic devices on glass with interdigitated geometry in Supplementary Table 2. The three devices are similar in geometry (~3 mm x 15 mm) and have interdigitated electrode structure, which makes the comparison of device performance straightforward. Several points stem from the Table: 1) The device demonstrated in this work operates at lower frequency. This is mainly due to thinner substrate (300 μm vs. 500 μm). 2) Device in this work needs lower $U_{rms}$ to achieve 1 μm displacement. This is mainly due to decreased gap between the fingers (3 μm vs. 10 μm, see Equation (2)) and lower substrate thickness. 3) Higher total capacitance in the current device is mainly due to a combination of smaller gap and finger width (5 μm vs. 10 μm). Most importantly, this device shows similar power consumption (35 mW) compared to the other two devices, which confirms its high quality.

Supplementary Table 2: **Comparison of piezoelectric thin-film haptic devices on glass with interdigitated geometry.** $f$ – resonant (operating) frequency; $U_{rms}$ – root mean square (rms) voltage at 1 μm deflection; $C_{device}$ – capacitance of the device; $P_{cons}$ – power consumption estimated as $P_{cons} = C f (U_{rms})^2$. In all three cases thickness of PZT was 1 μm and devices had similar geometries.

| Device | Glass | $f$ (kHz) | $U_{rms}$ (V) | $C_{device}$ (pF) | $P_{cons}$ (mW) |
|---|---|---|---|---|---|
| Spin-coated[4] | Fused silica | 73.0 | 43 | 240 | 32 |
| Inkjet-printed[11] | Fused silica | 63.3 | 46 | 230 | 31 |
| This work | AF32 | 40.2 | 34 | 760 | 35 |

**Supplementary Note 3.3: Electrical measurements**

Supplementary Figure 12 shows initial $P(\mathbf{E})$ and $J(\mathbf{E})$ loops for a single haptic actuator. Similarly pinched loops are observed as on fused silica (see Supplementary Figure 7). Note that this initially pinched behaviour does not influence haptic performance, as was demonstrated in inkjet-printed (RTA-processed) PZT films with similar behaviour[12]. Permittivity and losses values at zero field are 350 and 0.07, respectively.

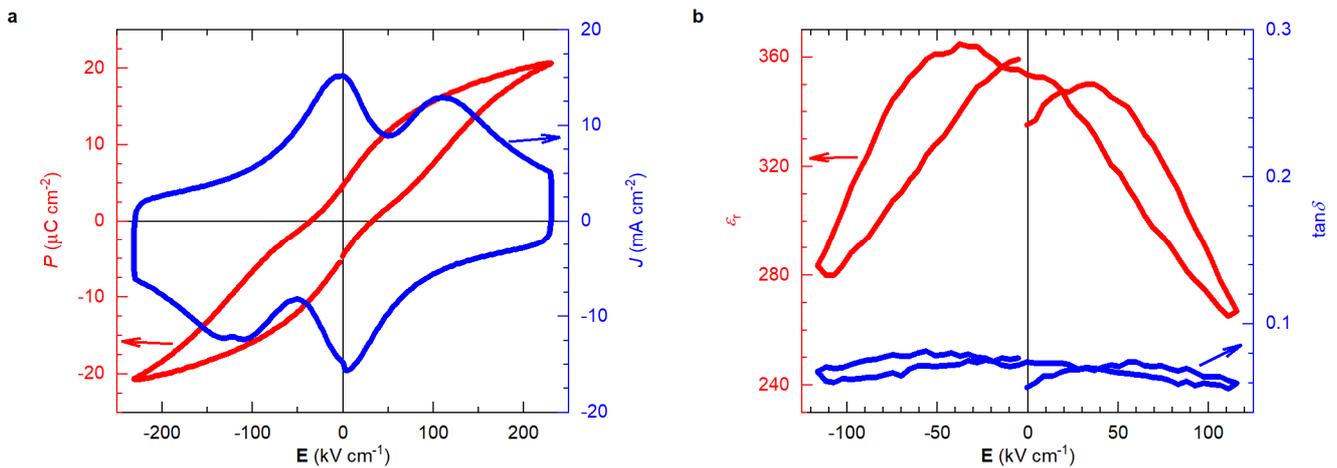

Supplementary Figure 12. **Ferroelectric and dielectric characterizations** of 1 μm thick PZT actuator on Schott AF32 glass. a) $P(\mathbf{E})$ and $J(\mathbf{E})$ loops of per actuator, measured at 100 Hz. b) corresponding $\varepsilon_r(\mathbf{E})$ and $\tan\delta(\mathbf{E})$ loops of the actuator, measured as functions of DC voltage with a probing AC signal of 0.5 V at 1 kHz.

**Supplementary Note 4: Flash lamp process for growing films on soda lime glass**

   **Supplementary Note 4.1: Two-step flash lamp annealing process**

The one-step process used for fused silica and AF32 glass is not suitable for growing PZT film on soda lime glass due to appearance of cracks. This can be attributed to the low thermal conductivity (1.0 W m$^{-1}$ K$^{-1}$) of the substrate, which leads to a slower rate of heat transfer and consequently a higher temperature at the interface between the film and glass.

To address this issue, we have developed a two-step process consisting of stages where either nucleation or growth is dominating. In the first step, pulses with higher power density are applied to induce the formation of nuclei within the film. This formation of nuclei reduces the activation energy required for the phase transition from an amorphous to a crystalline phase. In the second step, the phase where crystal growth is dominating, pulses with a lower power density are applied to grow the film at a lower temperature, thereby preventing the occurrence of cracks.

   **Supplementary Note 4.2: Phase composition characterization**

Supplementary Figure 13 displays GIXRD patterns of PZT films with thicknesses of 170 and 500 nm deposited on soda lime glass. The dominant reflections in both patterns correspond to the piezoelectrically active perovskite phase, suggesting that the FLA process is suitable for layer-by-layer preparation in solution processing. This is particularly significant for various applications. Although a weak reflection at approximately 29° indicates the presence of the secondary pyrochlore phase, its relative intensity is considerably lower than that of the perovskite reflections.

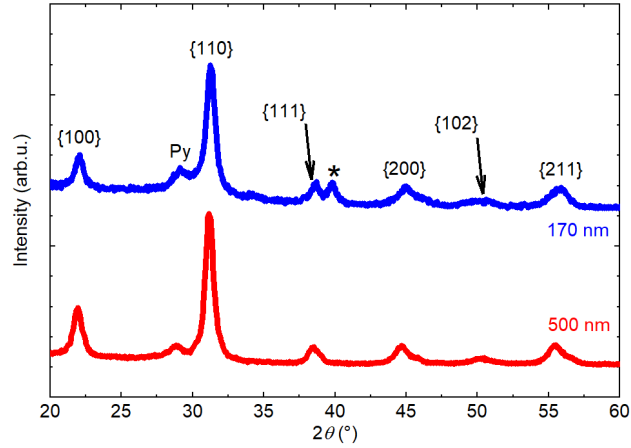

Supplementary Figure 13. **XRD study**. GIXRD patterns of 170 nm- and 500 nm-thick PZT films on soda lime glass. * denotes the signal of Pt IDEs on top of the PZT film. The PDFs No 01-070-4264 and 04-014-5162[3] have been used to identify the perovskite and pyrochlore phases, respectively.

**Supplementary Note 4.3: Electrical measurements**

The $P(\mathbf{E})$ and $J(\mathbf{E})$ loops of a 170 nm-thick PZT thin film on soda lime are shown in Supplementary Figure 14 along with its dielectric characterization.

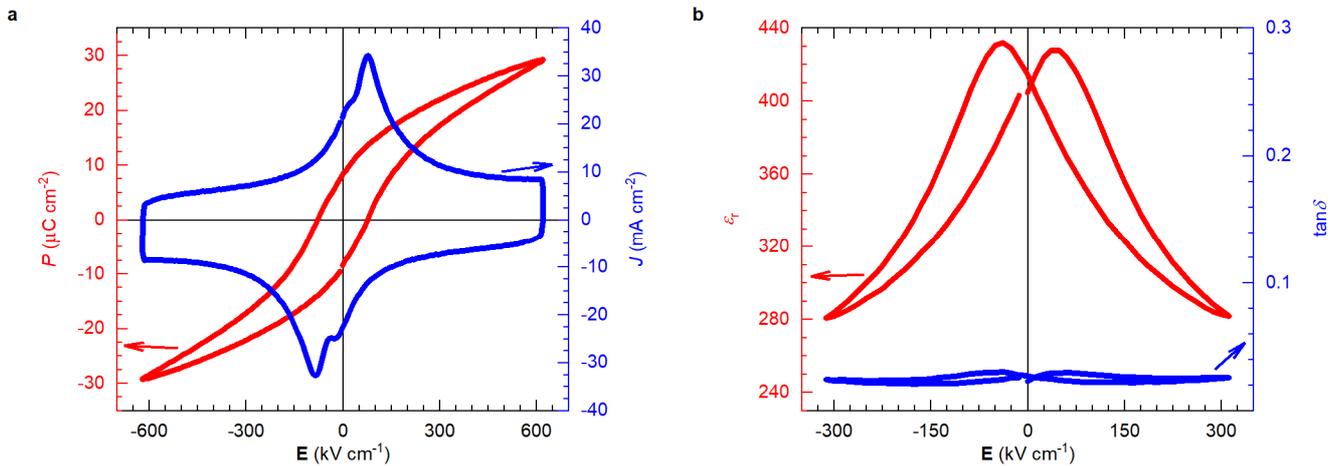

Supplementary Figure 14. **Ferroelectric and dielectric characterizations**. Ferroelectric and dielectric measurements of 170 nm-thin PZT on soda lime glass. a) $P(\mathbf{E})$ and $J(\mathbf{E})$ loops, measured at 100 Hz. b) $\varepsilon_r(\mathbf{E})$ and $\tan\delta(\mathbf{E})$ loops, measured at 1kHz. Small IDEs was used, whose parameters are 5 µm in digit width, 3 µm in interdigital gap, 50 pairs of digits and 370 µm in length of digits facing each other.

## Supplementary Note 5: Comparison with previous works

Supplementary Table 3: **Summary** of relevant points in the references previously reporting FLA treatment of PZT films and major advancement shown in this work.

| Reference | Relevant points in the reference | Major advancement in our work compared to reference |
|---|---|---|
| Yamakawa et al., Jpn. J. Appl. Phys., 41 2630 (2002)[13]. | <ul><li>FLA treatment of PZT thin films.</li><li>Ambient temperature between 300 and 500 °C.</li><li>Crystalline perovskite phase present prior FLA treatment.</li><li>No piezoelectric properties.</li></ul> | <ul><li>Ambient environment at room temperature.</li><li>Crystallization of completely amorphous initial films.</li><li>Demonstration of piezoelectric properties and a device.</li></ul> |
| Yao et al., J. Eur. Ceram. Soc., 40, 5396 (2020)[14]. | <ul><li>FLA treatment of PZT thin films.</li><li>No macroscopic electromechanical characterization (films too leaky).</li></ul> | <ul><li>Demonstration of macroscopic electromechanical properties and a device.</li></ul> |
| Palneedi et al., Adv. Mater., 2303553 (2023)[15]. | <ul><li>FLA sintering of crystalline PZT powders deposited on metglas (amorphous metal).</li></ul> | <ul><li>In-situ FLA crystallization of amorphous PZT thin films.</li></ul> |
| Ouyang et al., J. Am. Ceram. Soc., 99, 2569 (2016)[16]. | <ul><li>FLA sintering of crystalline PZT powders on stainless steel.</li><li>Non-saturated P-E loops.</li></ul> | <ul><li>In-situ FLA crystallization of amorphous PZT thin films.</li><li>Good ferroelectric properties.</li></ul> |
| Ouyang, PhD Thesis, Rochester Institute of Technology (2017)[17]. | <ul><li>FLA sintering crystalline PZT powders on stainless steel and PET substrates.</li><li>Non-saturated P-E loops.</li></ul> | <ul><li>In-situ FLA crystallization of amorphous PZT thin films.</li><li>Good ferroelectric properties.</li></ul> |
| Marotta, MSc Thesis, Rochester Institute of Technology (2019)[18]. | <ul><li>FLA treatment of printed PZT thin films.</li><li>No crystallization (XRD) reported.</li><li>No macroscopic electromechanical characterization.</li></ul> | <ul><li>In-situ FLA crystallization of amorphous PZT thin films.</li><li>Demonstration of macroscopic electromechanical properties.</li></ul> |

Two major points are stemming from the Table:

1) In all the previous reports with demonstrated macroscopic functional properties, FLA sintering was performed on already crystalline PZT powders. In this work, FLA crystallization, i.e., nucleation of perovskite grains and their growth, was performed from amorphous films (see Figure 2a of the main article). Perovskite formation is nucleation-controlled, with activations energies for nucleation and grain growth of 441 kJ mol$^{-1}$ and 112 kJ mol$^{-1}$, respectively[19]. In other words, the most energetically demanding process for perovskite formation is nucleation from the amorphous phase, which this study is the only one to demonstrate with FLA.

2) In the remaining reports, where they worked on FLA treatment of amorphous PZT films, macroscopic ferroelectric results could not be obtained. The only exception is the work of Yamakawa et al.[13] on sputtered PZT films. In that case, the FLA treatment was performed at elevated ambient temperatures and XRD reveals the presence of the perovskite phase already before FLA treatment (Figure 8 in the article). Therefore, this work is the only one that shows macroscopic ferroelectric results starting from fully amorphous films.

Supplementary Table 4: **FLA processing parameters taken from the literature and compared to the parameters used in this work.** Note that Ouyang's PhD thesis[17] is omitted as its results are summarized in the article in Journal of the American Ceramics Society[16]. Marotta's MSc thesis[18] is omitted from this analysis also, as little information on FLA process is given. *Energy per pulse in Yamakawa's work[13] is estimated from the current delivered to the Xe lamp when the voltage is applied. Real energy delivered to the sample is probably much lower. +Power per pulse (unless given) is estimated from the energy divided by pulse width. #In this work energy delivered to the sample was measured using bolometer.

| Reference | Energy per pulse (J cm$^{-2}$) | Pulse width (μs) | Power per pulse+ (kW cm$^{-2}$) | Number of pulses | Pulse frequency (Hz) |
|---|---|---|---|---|---|
| Yamakawa et al., Jpn. J. Appl. Phys., 41 2630 (2002)[13]. | 27* | 1000-1500 | 18-27 | up to 5 | not given |
| Yao et al., J. Eur. Ceram. Soc., 40, 5396 (2020)[14]. | not given | 250-500 | Up to 6.4 | up to 100 | not given |
| Palneedi et al., Adv. Mater., 2303553 (2023)[15]. | 1.7 – 7.4 | 250-1000 | 7 | up to 3 | 1 |
| Ouyang et al., J. Am. Ceram. Soc., 99, 2569 (2016)[16]. | 2.8 | 1300 | 2.2 | Up to 15 | 2 |
| This work | 3# | 130 | up to 23 | Up to 100 | 3 |

**Supplementary References**

1. Guillot, M. J., McCool, S. C. & Schroder, K. A. Simulating the thermal response of thin films during photonic curing. *ASME Int. Mech. Eng. Congr. Expo. Proc.* **7**, 19–27 (2013).

2. Piper, R. T., Daunis, T. B., Xu, W., Schroder, K. A. & Hsu, J. W. P. Photonic Curing of Nickel Oxide Transport Layer and Perovskite Active Layer for Flexible Perovskite Solar Cells: A Path Towards High-Throughput Manufacturing. *Front. Energy Res.* **9**, 1–12 (2021).

3. ICDD database PDF4+ v.19. (2019).

4. Glinsek, S. *et al.* Fully transparent friction-modulation haptic device based on piezoelectric thin Film. *Adv. Funct. Mater.* **30**, 2003539 (2020).

5. Nguyen, C. H. *et al.* Probing-models for interdigitated electrode systems with ferroelectric thin films. *J. Phys. D. Appl. Phys.* **51**, 175303 (2018).

6. Muralt, P. Recent progress in materials issues for piezoelectric MEMS. *J. Am. Ceram. Soc.* **91**, 1385–1396 (2008).

7. AF32 glass datasheet. https://www.pgo-online.com/intl/af32.html (2022).

8. Song, L. *et al.* Piezoelectric thick film for power-efficient haptic actuator. *Appl. Phys. Lett.* **121**, 212901 (2022).

9. Bernard, F., Casset, F., Danel, J. S., Chappaz, C. & Basrour, S. Characterization of a smartphone size haptic rendering system based on thin-film AlN actuators on glass substrates. *J. Micromechanics Microengineering* **26**, 84007 (2016).

10. Bernard, F. Conception, fabrication et caractérisation d'une dalle haptique à base de microactionneurs piézoélectriques. (Université Grenoble Alpes, 2016).

11. Hua, H., Chen, Y., Tao, Y., Qi, D. & Li, Y. A highly transparent haptic device with an extremely low driving voltage based on piezoelectric PZT films on glass. *Sensors Actuators A Phys.* **335**, 113396 (2022).

12. Glinsek, S. *et al.* Inkjet-printed piezoelectric thin films for transparent haptics. *Adv. Mater. Technol.* **7**, 2200147 (2022).

13. Yamakawa, K. *et al.* Novel Pb(Ti, Zr)O3 (PZT) crystallization technique using flash lamp for ferroelectric RAM (FeRAM) embedded LSIs and one transistor type FeRAM devices. *Japanese J. Appl. Physics, Part 1 Regul. Pap. Short Notes Rev. Pap.* **41**, 2630–2634 (2002).

14. Yao, Y. *et al.* Direct processing of PbZr0.53Ti0.47O3 films on glass and polymeric substrates. *J. Eur. Ceram. Soc.* **40**, 5369–5375 (2020).

15. Palneedi, H. *et al.* Intense pulsed light thermal treatment of Pb(Zr,Ti)O3/metglas heterostructured films resulting in extreme magnetoelectric coupling of over 20 V cm-1 O-1. *Adv. Mater.* 2303553 (2023) doi:10.1002/adma.202303553.

16. Ouyang, J., Cormier, D., Williams, S. A. & Borkholder, D. A. Photonic sintering of aerosol jet printed lead zirconate titanate (PZT) thick films. *J. Am. Ceram. Soc.* **99**, 2569–2577 (2016).

17. Ouyang, J. Enhanced piezoelectric performance of printed PZT films on low temperature substrates. (Rochester Institute of Technology, 2017).

18. Marotta, A. R. Printable thin-film sol-gel lead zirconate titanate (PZT) deposition using nanojet and inkjet printing methods. (Rochester Institute of Technology, 2019).

19. Chen, K. C. & Mackenzie, J. D. Crystallization kinetics of metallo-organics derived PZT thin film. *MRS Online Proc. Libr.* **180**, 663–668 (1990).